# Defect Engineered 2D $MoS_2$ Materials for ML-enabled Neurotransmitter SERS Detection

*Md Arifur R. Khan[1], Besan Khader[1], Nicholas Trainor[2], Chen Chen[3], Joan M Redwing[2,3], Slava V Rotkin[4,5], and Tetyana Ignatova[1]**

[1] - Department of Nanoscience, Joint School of Nanoscience and Nanoengineering, the University of North Carolina at Greensboro, Greensboro, NC, United States.
[2] - Department of Materials Science and Engineering, the Pennsylvania State University, University Park, PA 16802, United States

[3] - 2D Crystal Consortium Materials Innovation Platform, Materials Research Institute, the Pennsylvania State University, University Park, PA 16802, United States

[4] - Department of Engineering Science and Mechanics, the Pennsylvania State University, University Park, PA 16802, United States.

[5] - Department of Physics, the Pennsylvania State University, University Park, PA 16802, United States.

**Abstract.** An attachment of catechol-containing neurotransmitter molecules is demonstrated on defect-engineered two-dimensional $MoS_2$ platform, leading to activation of SERS due to molecular charge transfer. Mechanisms of neurotransmitters' bio-detection are discussed and Machine Learning methods are applied to distinguish spectra of structurally similar analytes. The SERS effect and selective docking of biomolecules are achieved through an optimized approach for defect engineering: namely, introducing the sulfur vacancies in $MoS_2$ monolayer films via soft plasma etching led to molecular attachment driven by catechol functional groups. The quality of the sensor material was controlled by Raman, photoluminescence, and XPS characterization, thus allowing for optimization of the process of defect formation and achieving sensing selectivity. The sensor material showed no response to serotonin, confirming the specificity of attachment/SERS due to S vacancies that regulate the strength of catechol-specific molecular adsorption. Defect-engineered $MoS_2$ has enabled SERS detection of dopamine and epinephrine down to the sub-nanomolar range ($5\times10^{-10}$ M), with strong calibration reliability ($R^2$ = 0.95 and 0.99 for pure samples). PCA-LDA achieved 100% accuracy in distinguishing dopamine and epinephrine, which establishes defect-engineered $MoS_2$ as a tunable, low-cost SERS platform for future sensing applications.

## 1. Introduction

Surface-enhanced Raman spectroscopy (SERS) is a powerful technique widely applied in healthcare, security, and environmental monitoring for the sensitive detection of target molecules by label-based and label-free methods. It is known that the performance of SERS devices is critically dependent on the proper design and careful preparation of the SERS-active substrate. [1–5] Two main models for SERS signal enhancement include an electromagnetic field hot spot generation (EM) and a chemical shift of energy resonances (CM). For metallic substrates, especially noble metals such as Au, Ag, etc., SERS enhancement is well-understood and primarily achieved through EM mechanism. A strong local electric field of a surface plasmon excitation is highly localized and depends on the size and shape of metal nano/micro structures at the surface.[6–13] Typically, SERS plasmonic substrates are functionalized for better analyte capture. Unlike EM model, for CM enhancement, which is typically pronounced in non-metallic materials, the charge transfer happens between the substrate and the target molecule.[14–18] This results in bringing the electronic transitions in resonance with the Raman excitation source and, thus, greatly increasing the SERS signal. While bulk/three-dimensional metal-based SERS substrates are extensively studied for their effectiveness in detecting various molecules, they have limitations in cost, stability, biocompatibility, and selectivity.[19] This has driven research interest in alternative SERS platforms, including two-dimensional (2D) materials. These studies gained momentum after the discovery of graphene-enhanced Raman scattering effect in the early 2010s.[20] [21] More recently, other 2D materials, [22–24] including transition metal dichalcogenides (TMDs) [25–28], have demonstrated strong SERS effect, later extended to hexagonal boron nitride [29,30], metal carbides [31], nano-compounds and van der Waals Fseroheterostructures.[32] 2D-SERS effect is often attributed to an enhanced molecular polarizability, stemming from the electronic coupling and charge transfer between the analyte and 2D substrate, which amplifies the Raman signal of the analyte.[33–35] The charge-transfer based enhancement was demonstrated in semiconductor systems, including heterostructures, defect-engineered metal-oxides [36–44], as well as semiconductor-metal hybrid materials.[45] In these studies, dye molecules like Rhodamine 6G were often used, which is typically in- or near-resonance conditions with the excitation source, leading to combined resonance Raman and substrate-induced SERS enhancement.[19,38,41,43,46–52] The resonant spectral overlap complicates the isolation of substrate-specific contributions, obscuring a precise identification of SERS mechanisms, and necessitates dedicated studies on CM enhancement. Indeed, the lack of understanding of the specific physical mechanism may prohibit rational design of future composite materials and hinder the practical implementations of 2D semiconducting SERS sensors.

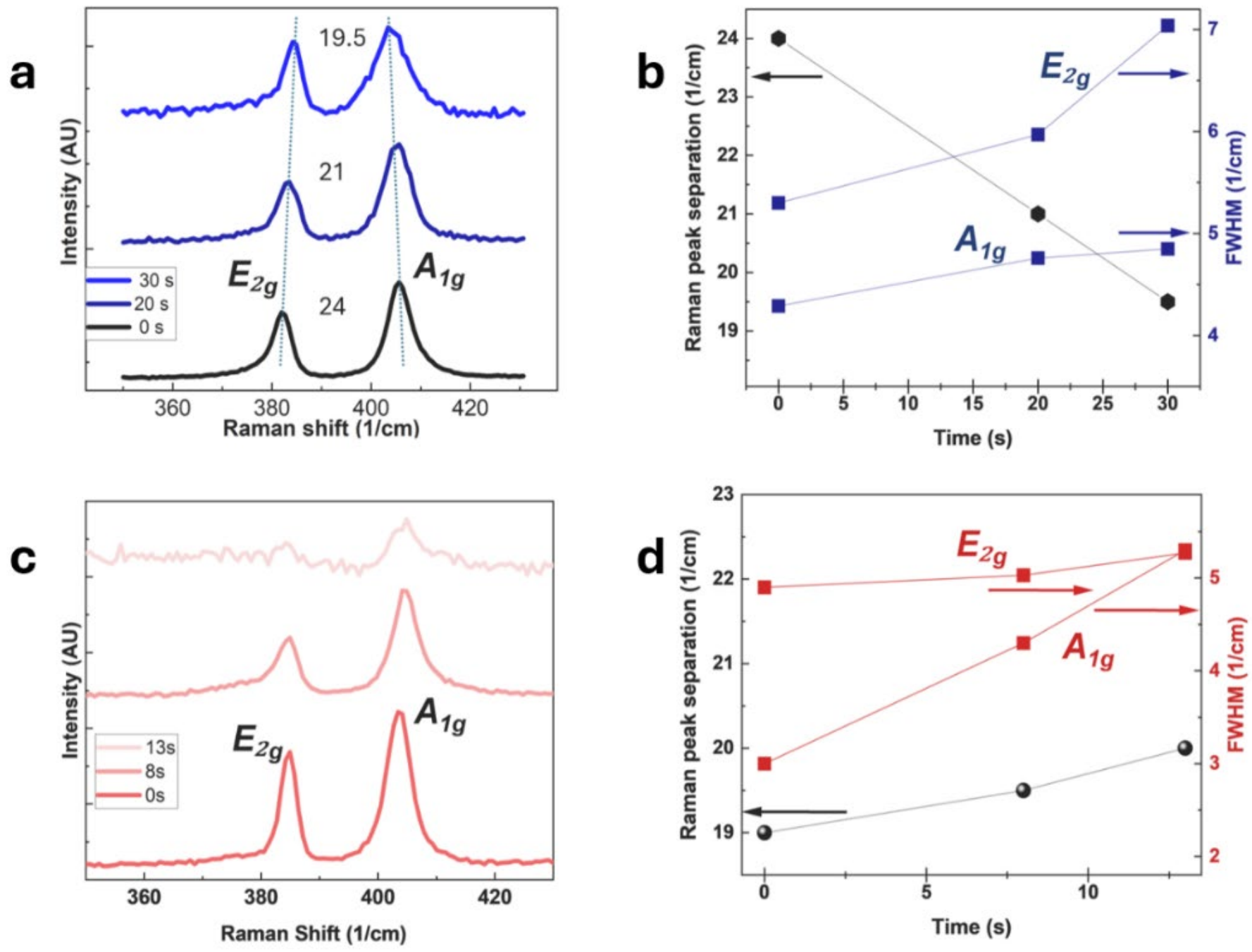

**Figure 1**. Raman characterization of multilayer $MoS_2$ after plasma treatment: reduction in the peak separation with increasing etching time (0 s, 20 s, and 30 s) (a); linear dependence of the $E_{2g}$ – $A_{1g}$ peak separation and broadening of Raman peaks on etching time (b). Raman characterization of monolayer $MoS_2$ after plasma treatment: reduction in intensity of the characteristic Raman peaks of monolayer $MoS_2$ with increasing etching time (0 s, 8 s, and 13 s) (c); peak separation and broadening dependence on etching time (d).

2D $MoS_2$, like other 2D materials, offers an outstanding SERS platform [19,53,54] due to its large specific surface area[55], tunable electronic structure[56,57], strong adsorption capabilities, and existence of charge-activated vacancy sites.[58–62] Importantly, the state-of-the-art synthetic methods are capable to produce large area $MoS_2$ of high crystalline quality and with controlled layer's number. However, SERS activity of bare $MoS_2$ is generally lower than that of metals, which limits its practical applications. Previous studies have demonstrated that structural imperfections, such as vacancies, dislocations, and grain boundaries, can enhance SERS activity of $MoS_2$ and increase interactions with a number of target molecules.[18,63,64] For instance, sulfur vacancies introduce localized states within the bandgap of $MoS_2$, altering its electronic structure, and creating external active sites, facilitating selective molecule adsorption.[17,65] Several methods, such as thermal etching, chemical reduction, and ion beam irradiation, have been used to introduce defects into $MoS_2$, each varying in complexity and effectiveness.[66–69] Notably, such defects lead to the electronic structure modulation which should also promote the charge transfer/separation to be discussed here in detail.

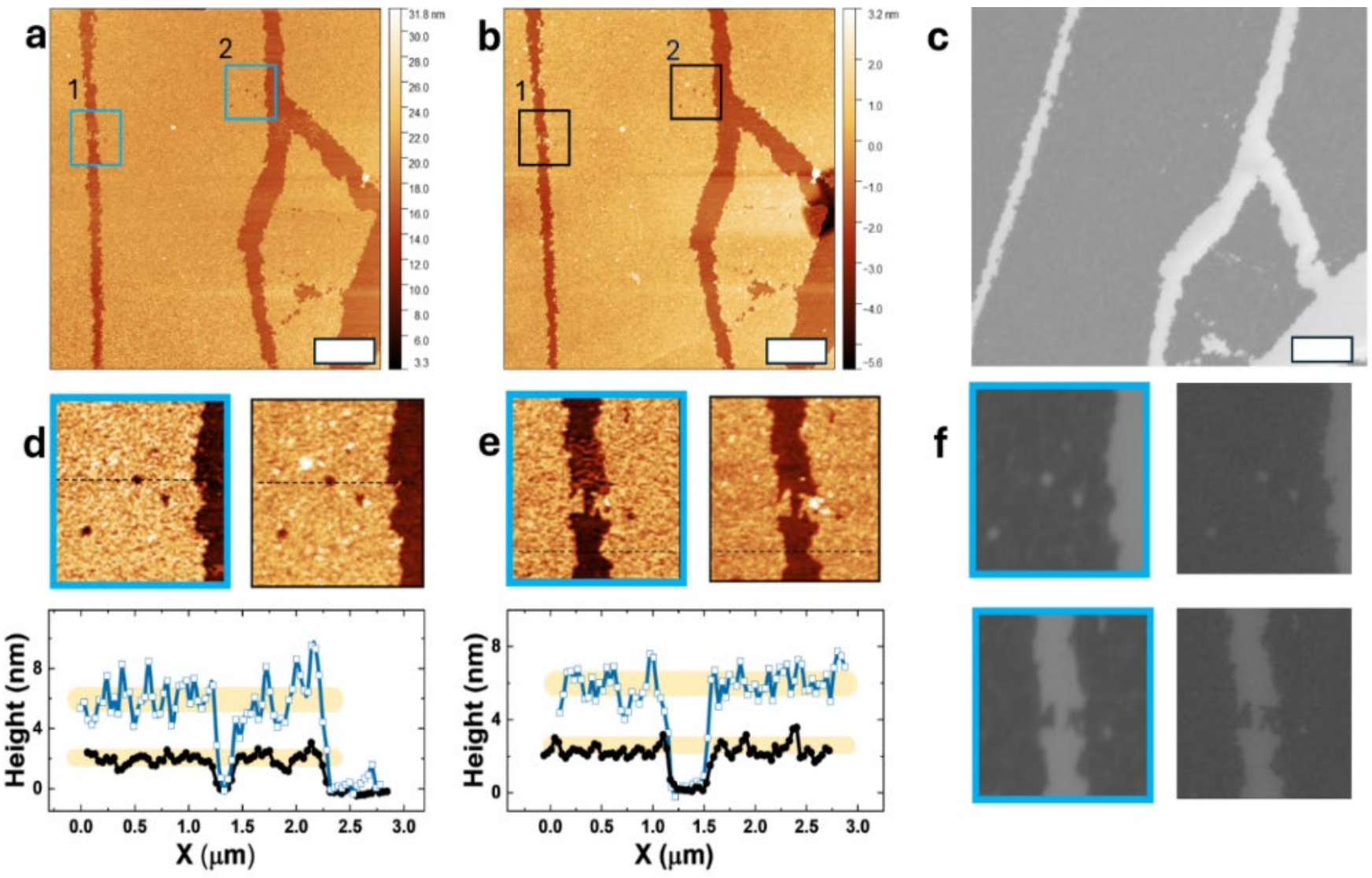

**Figure 2.** AFM images of multilayer $MoS_2$ sample before (a) and after (b) plasma etching; (c) SEM image of the same area before plasma etching; scale bar is 3 μm. The mag nified AFM and SEM images before plasma treatment (blue square) and after plasma treatment (black square) (d), (e), (f); AFM height profile taken along the same line before (blue) and after (black) plasma treatment.

In this study, we report a scalable approach for generating sulfur vacancies in 2D $MoS_2$ through reactive ion etching (RIE). Leveraging the number of sulfur vacancies allowed us to control the $MoS_2$ monolayer electronic properties. A thorough analysis of the defect-induced changes in electron subsystem of TMD material, resulting in the charge-transfer-induced enhancement of the SERS activity, suggests a strategy for designing defect-engineered SERS platforms for the label-free neurotransmitter detection.

Dopamine (DA), epinephrine (EP), and serotonin or 5-hydroxytryptamine (5-HT) are three prominent neurotransmitters responsible for critical physiological functions in the human body, ranging from motor control and stress response to emotional regulation.[70,71] The molecules of DA and EP contain catechol functional groups, whereas 5-HT does not. Removing sulfur atoms via selective plasma etching and exposing under-coordinated Mo atoms opens possibility of forming chelation bonds for analytes with catechol groups, similar approach used for SERS detection of DA with functionalized plasmonic particles.[72] In this work, DA and EP label-free SERS detection under controlled conditions via targeted chemical binding is demonstrated, using 5-HT as a negative reference. The catechol-Mo complexation gives rise to the resonant charge transfer between TMD and the analyte molecule. It corroborates the CM mechanism of SERS on plasma-treated $MoS_2$, where the latter acts as a selective molecular recognition platform for analytes containing catechol groups, which minimizes non-specific adsorption and improves signal-to-noise ratio. While established analytical techniques are commercially available, the proposed defect-engineered $MoS_2$ SERS platform offers the potential advantages of rapid, label-free, ultrasensitive, and potentially portable molecular detection with minimal sample preparation, which may be beneficial for future point-of-care and real-time sensing applications.

We demonstrate enhanced SERS sensitivity of defect-engineered $MoS_2$ with a limit of detection (LoD) as low as $0.5\times10^{-10}$ M for DA and EP, under controlled conditions. To isolate the intrinsic role of S vacancies in charge-transfer-induced SERS enhancement, measurements were intentionally performed using pure analytes in DI water, minimizing interference from competing species. The charge-transfer mechanism responsible for SERS and the improved sensitivity of the TMD platform are due to the specific interactions of sulfur vacancies with DA/EP, which is also validated by negligible SERS response for 5-HT, lacking the catechol functional group. These findings highlight the potential of defect engineered 2D $MoS_2$ as a cost-effective and tunable SERS platform for biomedical applications and propose a pathway for use of 2D semiconductors in detecting a large range of biologically relevant molecules.

## 2. Results and Discussion

### 2.1. Engineering Properties of 2D $MoS_2$ by Reactive Ion Etching

Using a soft plasma etching to modify 2D $MoS_2$ films is a known technique.[19,73–78] In order to achieve a balance between defect formation and retaining $MoS_2$ material, resulting in the best SERS activity, several parameters have been optimized in this study, such as etchant/buffer gas selection, gas flow rate, gas chamber pressure, power, and time.

| Power, W | Pressure, mTor | $SF_6$ Flow Rate, sccm | Ar Flow Rate, sccm | Time, s | Outcomes |
|---|---|---|---|---|---|
| Multilayer $MoS_2$ film | | | | | |
| 150 | 100 | 25 | 0 | 60 | Complete material's removal |
| 30 | 100 | 25 | 6 | 20 | ttwo layers removal |
| 30 | 100 | 25 | 6 | 30 | Thinning to monolayer |
| Monolayer $MoS_2$ film | | | | | |
| 10 | 100 | 25 | 6 | 8 | Introducing defects |
| 10 | 100 | 25 | 6 | 13 | Material's etching |
| 10 | 100 | 25 | 6 | 15 | Complete material's removal |

*Table 1 Summary of plasma parameters optimization and outcomes*

Selection of plasma gas is of utmost importance to alter material's properties while controlling the etching rate. $SF_6$ was chosen as a primary plasma gas source, with $F^-$ radicals interacting with $MoS_2$ in a two-step process leaving behind desired concentration of sulfur vacancies (defected TMD is denoted as $MoS_{2-n}$):

(1) $n\,SF_6 \rightarrow nSF_{6-n} + nF^-$

(2) $2nF^- + nMoS_2 \rightarrow nMoS_{2-n} + nSF_2\ \uparrow$

However, exposed Mo atoms are subject to etching as well:

(3) $6nF^- + nMoS_2 \rightarrow nMoF_6 \uparrow + 2nS$

Thus, this plasma etching is anticipated to lead to a layer-by-layer removal of the whole TMD material, only stopping when the entire $MoS_2$ film is eliminated.

During the optimization trial, aimed to control etching rate, the plasma power of 150 W and the etching duration of 60 seconds were selected, which caused a quick, complete removal of the $MoS_2$ film as confirmed by optical microscopy and by absence of the $MoS_2$ characteristic Raman peaks. To achieve selective removal of a desired layer number, we adjusted the etching parameters: while maintaining the same gas pressure at 100 mTorr and the flow rate at 25 sccm, we significantly reduced the power to 30 W and introduced a gas mixture of $SF_6$ and Ar in a 80:20 ratio, vs. use of the sole $SF_6$ in the initial experiment. The inclusion of buffer/non-reactive Ar fraction was designed to reduce the density of highly reactive $F^-$ ion radicals generated in the plasma, although it may cause a sputtering effect by energized Ar ions. At this gas ratio (80:20), we speculate that the time-dependent chemical reactions of $F^-$ ions dominate over Ar sputtering, thus resulting in controlled, layer-by-layer etching of $MoS_2$, as confirmed by Raman characterization detailed next. 2D $MoS_2$ exhibits two characteristic Raman peaks, corresponding to the in-plane vibration ($E_{2g}$) and the out-of-plane vibration ($A_{1g}$), typically observed at 384 $cm^{-1}$ and 404 $cm^{-1}$, respectively.[19] The peak position difference is indicative of the number of layers. For monolayer $MoS_2$, the difference is 19-20 $cm^{-1}$, increasing to 21 $cm^{-1}$ for bilayers, and even greater for the multilayers.[56,79–81] **Figure 1a** shows that the pristine $MoS_2$ sample has a peak separation of 24 $cm^{-1}$, confirming its multilayer structure. After plasma etching for 20 and 30 seconds, the peak separation is reduced to 21 $cm^{-1}$ and 19.5 $cm^{-1}$, respectively, indicating successful layer-by-layer removal of $MoS_2$. The linear dependency of the layer number dependent Raman shift on the etching time, shown in Figure 1b, supports the selectivity of the etching procedure. Moreover, the broadening of the Raman peaks (as shown in Figure 1b) after 30 seconds of etching suggests the intense introduction of defects into the 2D $MoS_2$ structure, followed by complete material removal upon further increase of etching time.

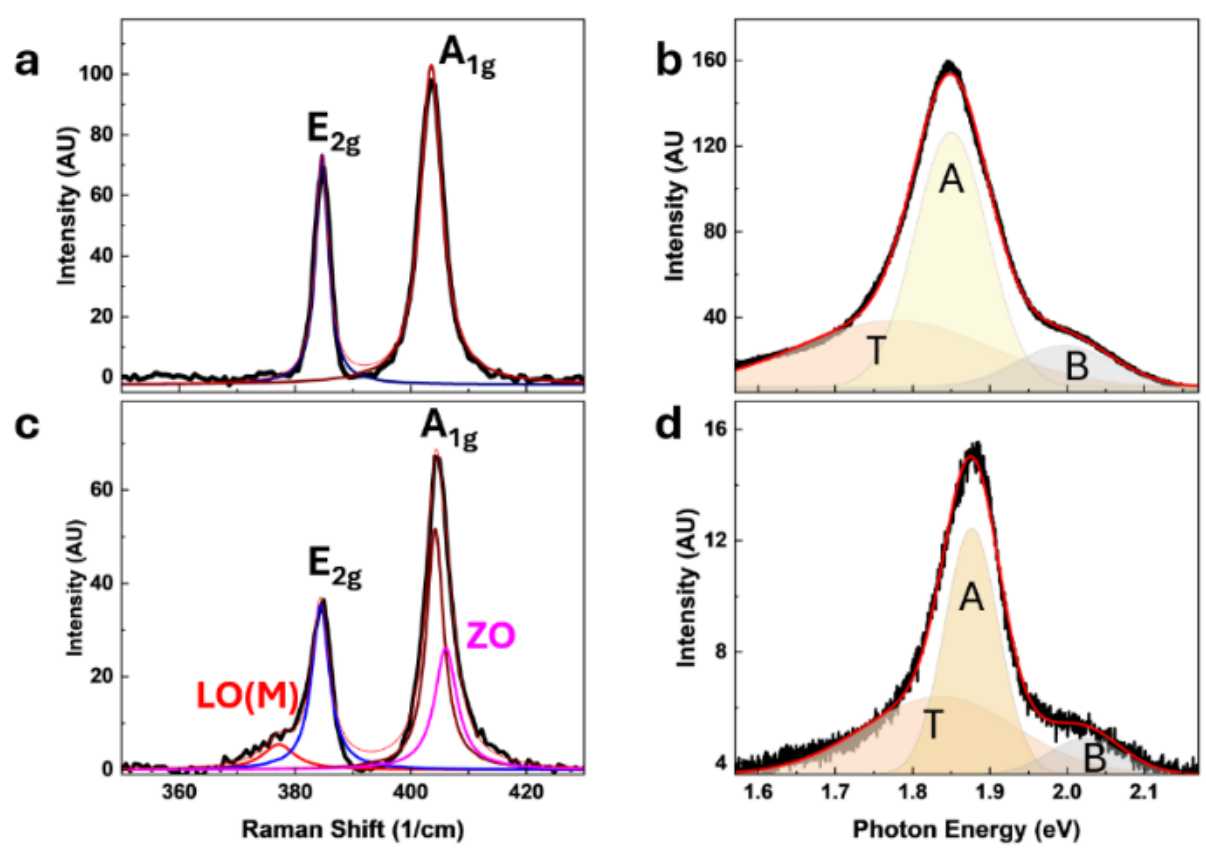

**Figure 3**. Raman spectra of $MoS_2$ before (a) and after (c) plasma etching. PL spectra of $MoS_2$ before (b) and after (d) plasma etching.

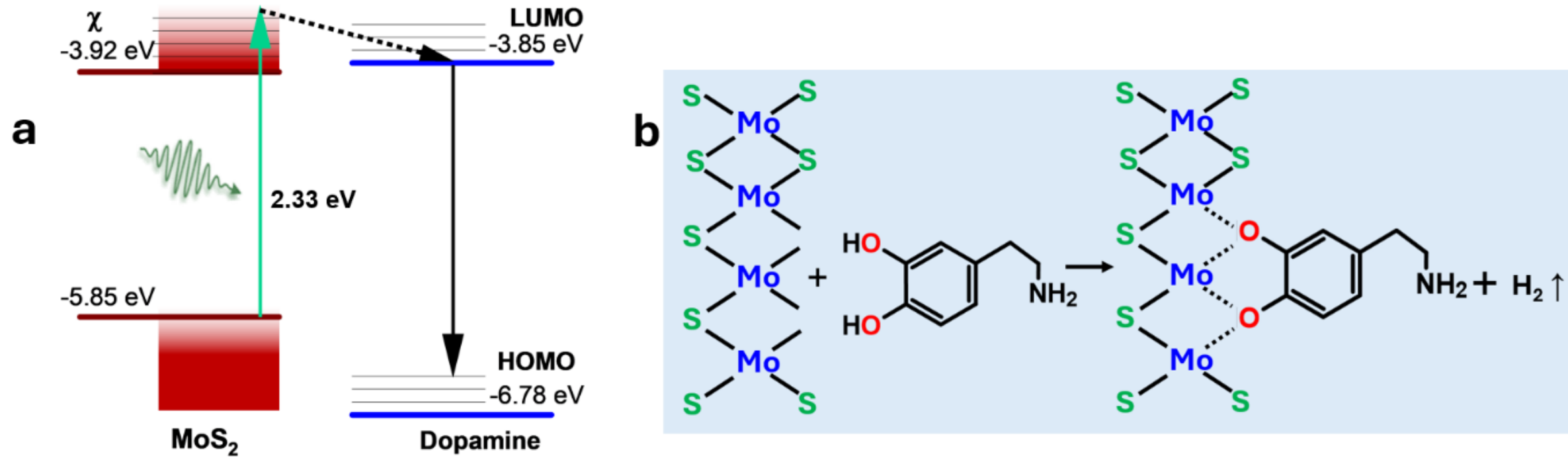

**Figure 4.** Schematic illustration of charge transfer mechanism between electronic bands of plasma-treated $MoS_2$ and dopamine orbitals (a). Schematic illustration of chemical interaction between dopamine and plasma-treated $MoS_2$ (b). The $E_v$ and Ec energy levels of $MoS_2$[86], as well as the HOMO-LUMO energy levels of dopamine[87], were adopted from previously reported studies and are shown here for illustrative comparison.

Final tuning of etching parameters for engineering of a monolayer $MoS_2$ film has been achieved on the next step: by reducing the plasma power to 10 W, while optimizing the etching time (0 to 13 seconds), we were able to introduce defects, sulfur vacancies, as desired, while preserving most of the layer material. Indeed, the reduction in Raman peak intensity and the $E_{2g}$ – $A_{1g}$ peak separation as a function of etching time are shown in Figure 1c, 1d (and Figure S1a, S1b, S1c), along with the peak broadening and tail formation. The $E_{2g}$ – $A_{1g}$ peak separation did not show an increase of more than 1 $cm^{-1}$. After 13 seconds of exposure, the Raman peaks were significantly diminished, indicating substantial modification of the $MoS_2$ crystal structure. We emphasize that, upon carefully optimized etching conditions, the monolayer film contingency is completely preserved – not even a slight edge etching was observed, as confirmed by thorough atomic force microscopy (AFM) and scanning electron microscopy (SEM) imaging, as shown in **Figure 2**.

A summary of the process optimization and its outcomes is presented in Table 1. Ultimately, for controllable creation of the sulfur defects in the monolayer $MoS_2$, we select a plasma gas mixture 80%: 20% ($SF_6$ : Ar), at 10 W power and 8-second exposure time (Table 1, Row 5).

The detailed surface morphology of the 3-layer $MoS_2$ sample was analyzed using AFM (Figure 2a, 2b) and SEM (Figure 2c, Figure S2) before and after plasma etching in the same region. AFM and SEM analysis revealed no evidence of surface degradation, such as appearance of additional holes, fractures, or structural discontinuities resulting from the plasma treatment. Furthermore, even the shape of the $MoS_2$ film boundary is preserved, including the smallest islands near the film edge - compare magnified blue square regions 1, 2 before plasma and black square regions 1, 2 after plasma etching shown on Figure 2d, 2e, 2f – which suggests that the etching process is slow enough to control the process, and has a vertical direction, such that the edge/horizontal etching is essentially prohibited under the optimized conditions. While the 2D film continuity is preserved, the morphological analysis via AFM showed that the film thickness was reduced by approximately 1 monolayer, reaching bilayer thickness (Figure 2d and 2e). Additionally, the reduced surface roughness observed after etching indicates that the $MoS_2$ surface has been efficiently cleaned through the removal of weakly bound molecular species and adsorbed impurities.

Detailed analysis of the Raman spectra was conducted for both the pristine monolayer $MoS_2$ and the plasma-treated (at optimized parameters) $MoS_2$. As expected, the pristine $MoS_2$ exhibits two prominent peaks at 385 $cm^{-1}$ and 404 $cm^{-1}$. The peaks' shape and their full width at half-maximum (FWHM) indicate a highly ordered crystalline structure with minimal defects (**Figure 3a**). The Raman spectrum obtained after plasma etching, reveals the appearance of defect-induced Raman bands leading to peak broadening and tail formation (Figure 3c). The new satellite peaks observed at ~377 $cm^{-1}$ and ~406 $cm^{-1}$ correspond to the defect-induced longitudinal optical (LO) and out-of-plane optical (ZO) bands at the M point of $MoS_2$, respectively, and according to previous reports, can be associated with the formation of sulfur vacancies. [82,83] The modification of Raman spectra in the etched sample is corroborated by photoluminescence (PL) measurements.

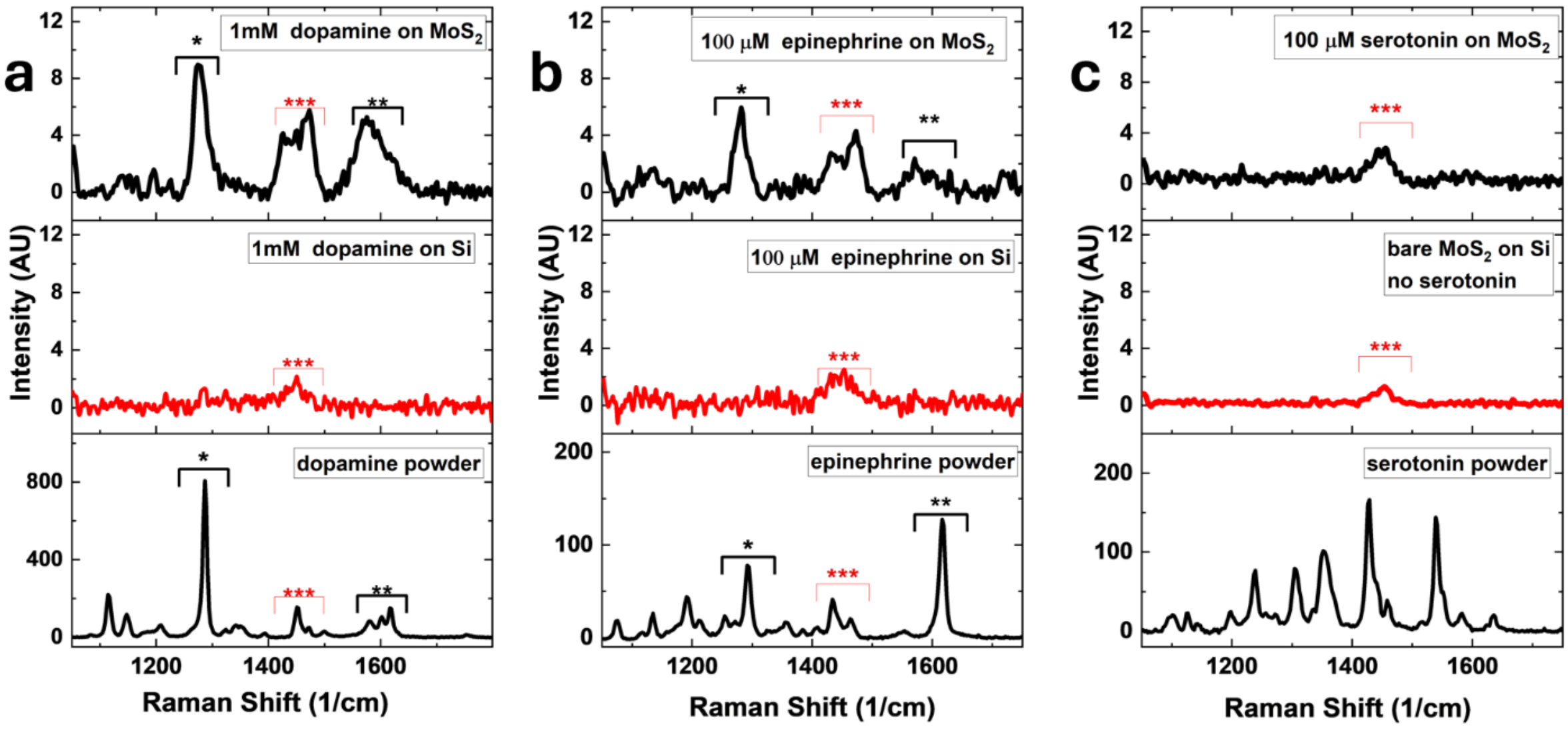

**Figure 5.** Raman spectra of the neurotransmitter analytes on plasma-treated $MoS_2$ substrate in the top row; the (negative) reference in the middle row; the reference (powder samples) in the bottom row. SERS spectra are shown for: (a) 1 mM dopamine, (b) 100 µM epinephrine, (c) 100 µM serotonin.

In the pristine $MoS_2$, three primary PL features are typically observed: the prominent A exciton peak between 1.8 and 1.9 eV, the less intense B exciton peak around 2.0 eV, and a negatively charged exciton (trion) below the A-exciton position (Figure 3b, peaks are fitted with Gaussian model).[53,84,85] When $MoS_2$ is subjected to plasma etching, defects drastically change the PL line shape. Firstly, the defects may introduce localized states within the bandgap, resulting in photocarrier trapping, which influences the dynamics of excitons and trions, as well as the concentration and mobility of free charge carriers. Additionally, these traps increase the non-radiative recombination rates. After plasma etching, the PL intensity of the same sample was significantly reduced, as seen in Figure 3d (and in Figure S1d).

Previous studies have demonstrated that the PL intensity of $MoS_2$ has a strong dependence on (p- or n-) doping level.[56,57] In our experiment, assuming that the observed reduction in PL intensity is an indirect confirmation of the formation of S vacancy sites, the n-doping should increase along with the defect concentration. This should reduce the exciton population via the non-radiative channels as well as promote the formation of trions. In the PL of the plasma-treated $MoS_2$ shown in Figure 3d, all peaks are significantly reduced in intensity, while the trion to A-exciton ratio increases from 0.7 in the pristine sample to 0.9 in the treated sample. The calculated S defect density was increased to $5.6 \times 10^{14}$ $cm^{-2}$ in the plasma-treated sample compared to the pristine sample, with details provided in the supplementary information (SI) along with XPS results (Figure S3).

**Figure 5.** Schematic illustration of charge transfer mechanism between electronic bands of plasma-treated $MoS_2$ and dopamine orbitals (a). Schematic illustration of chemical interaction between dopamine and plasma-treated $MoS_2$ (b). The $E_v$ and Ec energy levels of $MoS_2$[86], as well as the HOMO-LUMO energy levels of dopamine[87], were adopted from previously reported studies and are shown here for illustrative comparison.

Upon laser illumination, electrons are excited from the valence band (VB) at $E_v$~-5.85 eV to above the conduction band (CB) edge at $E_c$~-3.92 eV, promoting charge transfer from $MoS_2$ hot electrons to the LUMO state of the DA molecule at ~-3.85 eV, possibly facilitated by Herzberg–Teller vibronic coupling. This mechanism, shown in **Figure 4**, corroborates with the model of SERS enhancement via charge transfer in the previous studies.[33,88–91]

### 2.2. 2D MOS₂ SERS Platform for Neurotransmitter Detection

Molecular structure of DA, EP, and (5-HT) contains a large number of similar functional groups: hydrocarbons of aliphatic chains (-CH); methyl group ($CH_3$) for EP; amine group ($-NH_2$) in DA; hydroxyl group (-OH) for EP/5-HT; notably, DA and EP possess the same catechol group, lacking in 5-HT. They are present in biological systems within the nanomolar to micromolar range, depending on the physiological fluid (e.g., blood, urine, or saliva)[70,71,92]. Raman spectra of powder DA, EP, and 5-HT are shown in Figure S6, and their vibrational assignments of major Raman peaks are described in SI (Table S1).

Upon incubation of the plasma-treated $MoS_2$ in a 1 mM DA solution (**Figure 5a**) or in a 100 µM EP

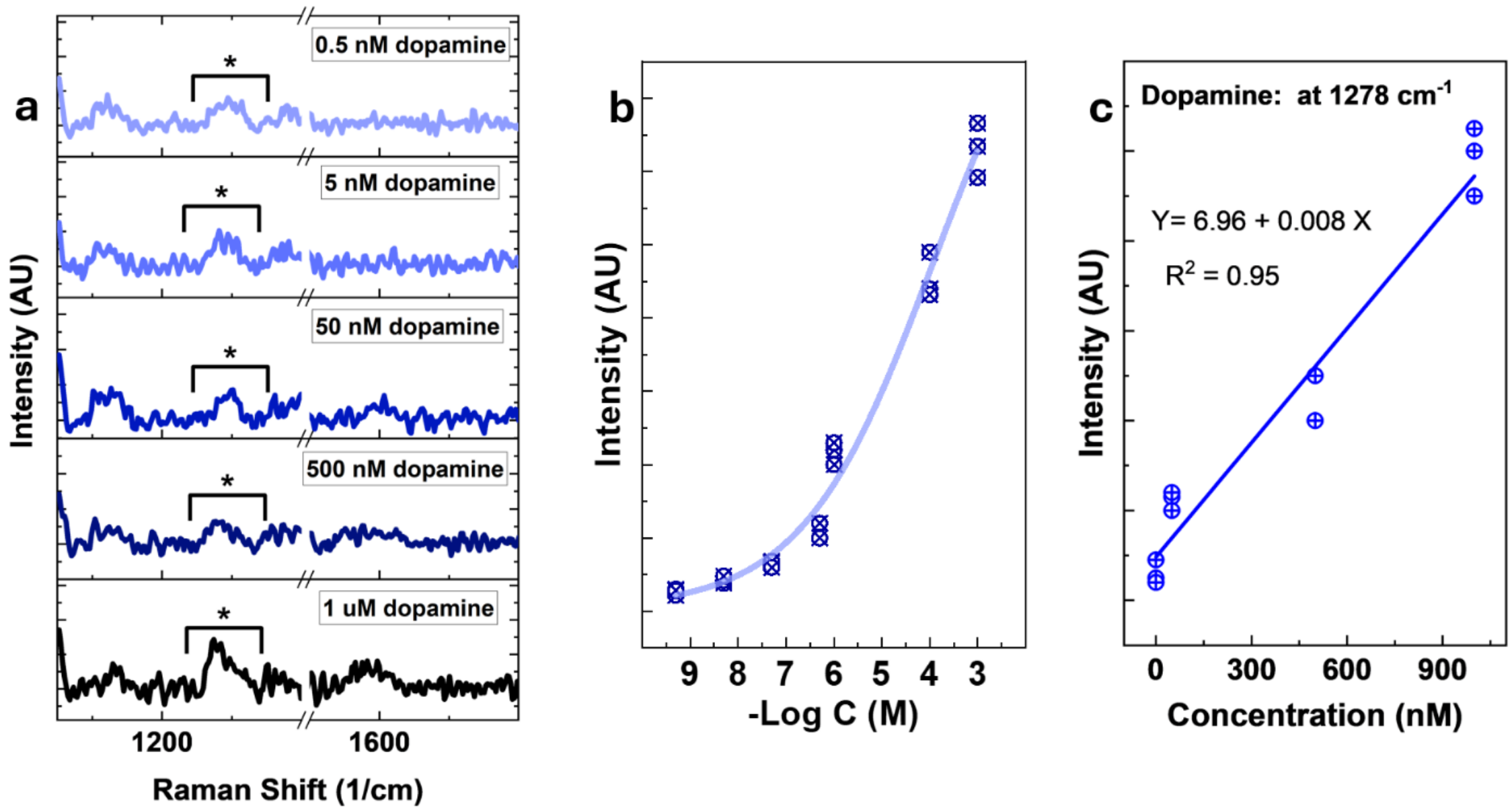

**Figure 6.** (a) SERS spectra of DA down to sub-nanomolar concentrations (10-6 to 10-10 M). (b) Plot of DA SERS intensity at 1278 cm⁻¹ (normalized to Si Raman peak at 520 cm⁻¹) vs. –log C (M) across concentrations ranging from 10-3 to 10-10 M. (c) Calibration curve showing the linear range for DA detection (0.5 nM to 1 µM) at the SERS band of 1278 cm⁻¹.

solution (Figure 5b), several characteristic peaks can be detected in the spectral range 1100 cm⁻¹ to 1700 cm⁻¹. To correct for the substrate effects, a peak at 1450 cm⁻¹ will be excluded from further analysis. Indeed, this feature is present in all samples, including the negative control and reference samples (bare $SiO_2$/Si substrate, and the plasma-treated $MoS_2$ incubated in deionized (DI) water w/o analyte). In the past, a similar Raman band was assigned to the 3TO overtone mode of Si[93]. Besides this background mode, new spectral features are observed that can be attributed to molecule-specific C–O and C–C vibrations, at 1278 cm⁻¹ and 1580 cm⁻¹ for DA, and at 1284 cm⁻¹ and 1570 cm⁻¹ for EP, respectively. A peak shift of up to 5-13 cm⁻¹ is noted when comparing the SERS spectra with the reference Raman spectra of the powder sample, which is normal for SERS measurements.[94–98] In contrast to DA and EP, no characteristic 5-HT peaks were detectable for the 5-HT solution at 100 µM, as shown in Figure 5c. This observation correlates with the absence of a catechol group, which may prevent serotonin from binding to S defects in $MoS_2$.

An important parameter for characterizing the performance of a SERS-active platform is the limit of detection (LoD), defined here as the lowest concentration at which characteristic Raman bands remain consistently distinguishable above background under controlled conditions. To measure SERS response to the DA analyte, SERS band at approximately 1278 cm⁻¹ was chosen, which is consistently present across all diluted conditions, ranging from 1 µM to 0.5 nM (**Figure 6a**). An additional peak around 1580 cm⁻¹ was detectable down to a concentration of 50 nM DA solution (**Figure S7**). Figure 6b shows the increase in intensity of the 1278 cm⁻¹ peak with concentration, based on multiple spectra taken at different locations. The calibration curve, shown in Figure 6c, demonstrates a strong linear relationship in the concentration range of 0.5 nM to 1 µM dopamine, with a correlation coefficient ($R^2$) of 0.95.

A similar analysis was performed for EP, which also contains a catechol functional group. For diluted EP solutions, the peak at 1284 cm⁻¹ remains prominent down to a concentration of 0.5 nM (**Figure 7a**). Additionally, peak at 1570 cm⁻¹ observed even in the diluted samples. Figure 7b shows the increase in intensity of the 1284 cm⁻¹ peak with concentration, based on multiple spectra taken at different locations. The characteristic peak at 1284 cm⁻¹ was used to construct a calibration curve (Figure 7c), showing a strong linear relationship in the concentration range of 0.5 nM to 1 µM epinephrine. The correlation coefficient ($R^2$) was found to be 0.99. These results confirm that the enhanced SERS response is governed by defect-mediated molecular interactions.

### 2.3. Machine Learning–Based Multivariate Analysis

Although dopamine and epinephrine exhibit catechol-mediated SERS activation on defect-engineered $MoS_2$, their spectra contain closely spaced and overlapping features that hinder direct visual differentiation. To resolve these subtle spectral variations and assess spectral consistency,

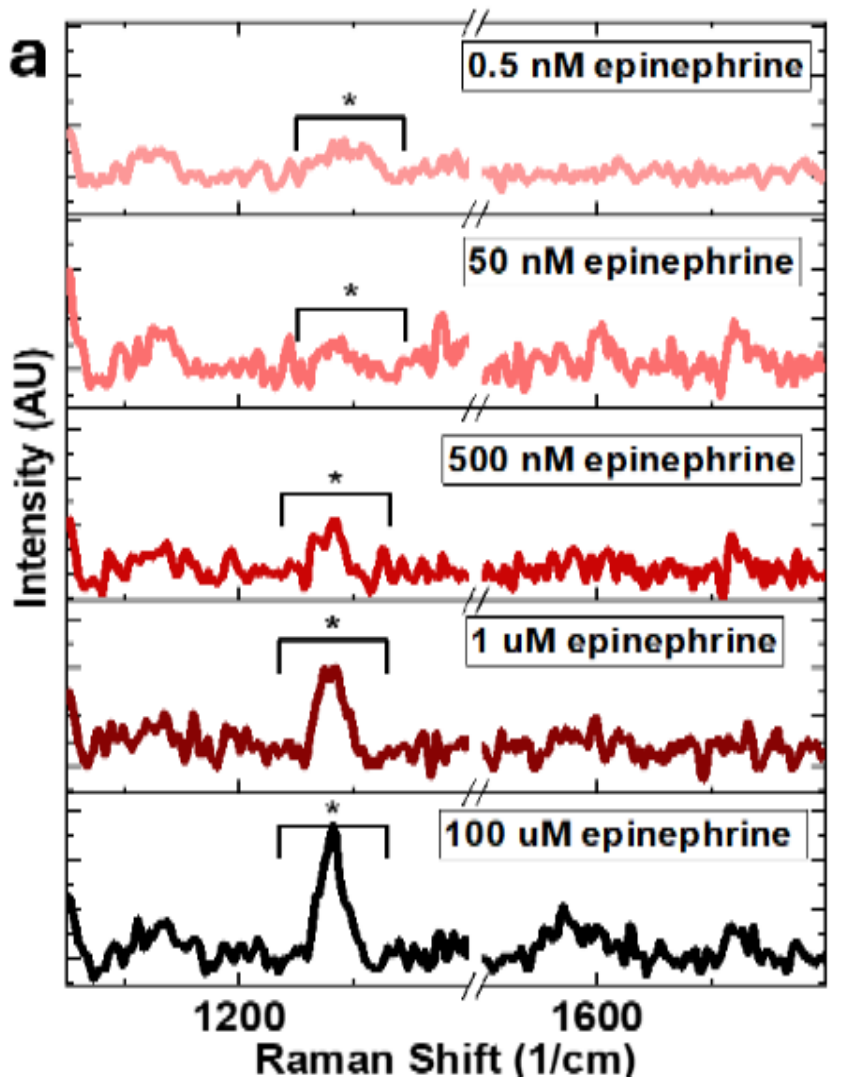

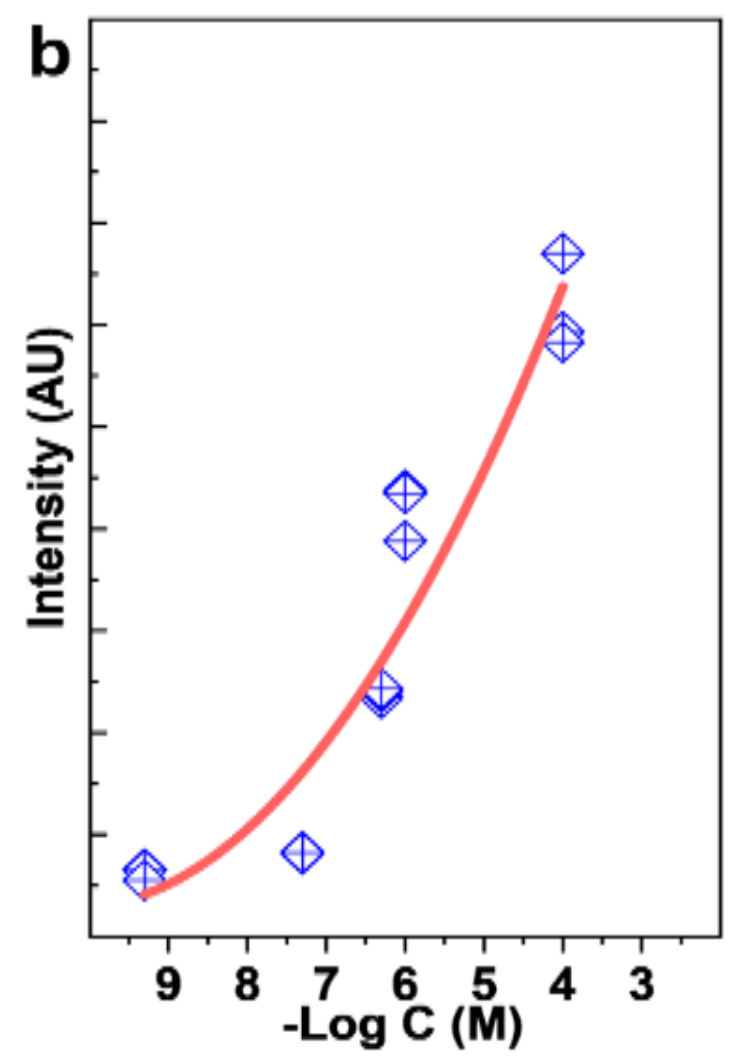

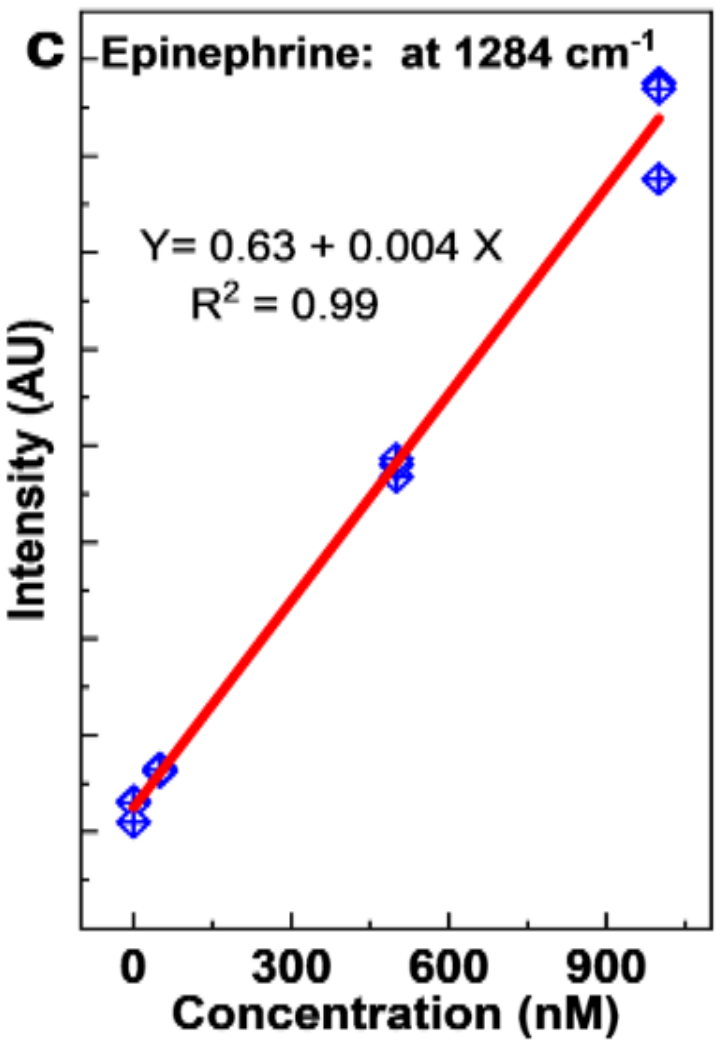

**Figure 7.** (a) SERS spectra of EP down to sub-nanomolar concentrations ($10^{-4}$ to $10^{-10}$ M). (b) Plot of EP SERS intensity at 1284 cm⁻¹ (normalized to Si Raman peak at 520 cm⁻¹) vs. –log C (M) across concentrations ranging from 10-4 to 10-10 M. (c) Calibration curve showing the linear range for EP detection (0.5 nM to 1 µM) at the SERS 1284 cm⁻¹.

principal component analysis (PCA) followed by linear discriminant analysis (LDA) was employed.

Principal Component Analysis (PCA) is a powerful statistical tool for analyzing complex datasets in Raman spectroscopy, where spectral features can vary subtly between analytes.[99–103] Appropriate preprocessing of spectral data is crucial for achieving consistency and minimizing variations caused by experimental fluctuations, such as laser instability, focus depth, and sample heterogeneity. To address these issues, we normalized all background-subtracted spectra against the silicon peak (~ 520 cm⁻¹), which served as an internal standard for laser intensity calibration. We then applied a first-derivative transformation combined with the Savitzky–Golay (SG) smoothing method. This technique enhances subtle spectral features while also decreasing baseline drifts and fluorescence background, thus improving both the signal-to-noise ratio and spectral resolution. Finally, we employed vector normalization to compensate for overall intensity differences between spectra, thereby facilitating direct comparisons across samples. Together, these preprocessing steps are crucial for revealing hidden spectral features and enabling machine learning models to focus on relevant spectral intensities, peak shapes, and relative distributions, thereby enhancing the sensitivity and reliability of the analysis. We excluded the Raman spectral region below 470 cm⁻¹ to prevent contributions from $MoS_2$-related peaks in the subsequent analysis.

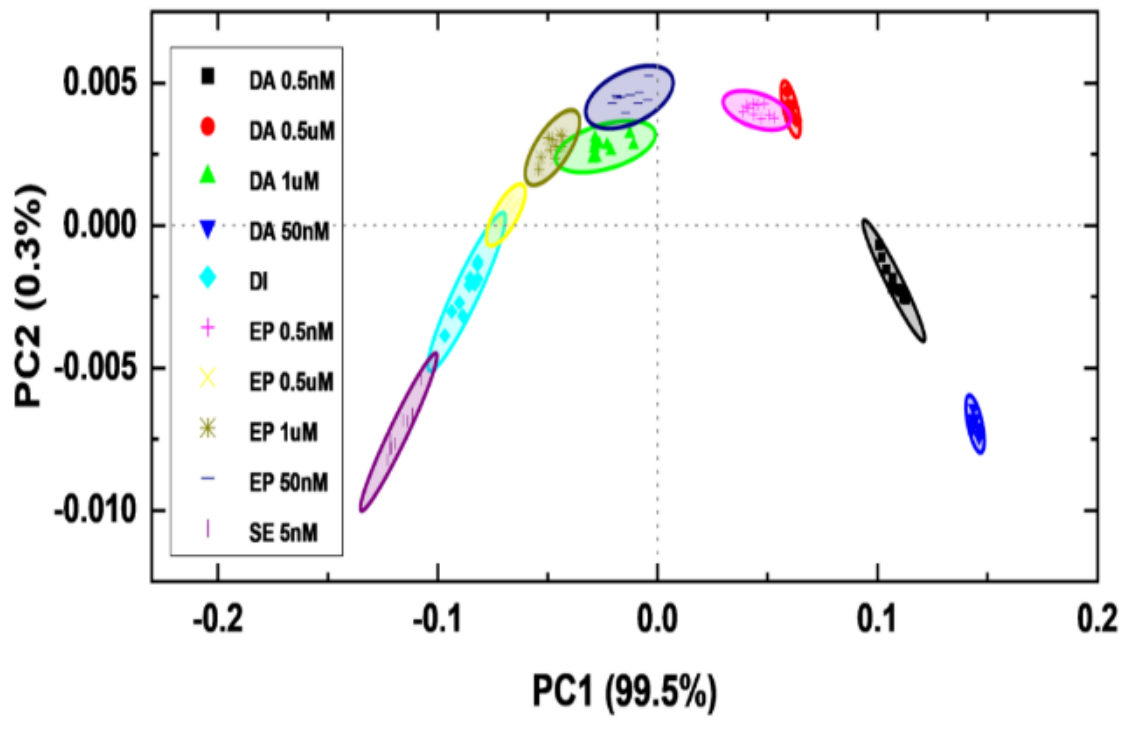

**Figure 8.** Multivariate analysis of SERS spectra for classification of analytes. PCA score plot illustrating spectral clustering based on the first two principal components, with 95% confidence ellipses highlighting class separation.

Following preprocessing, we used principal component analysis (PCA) to reduce the dimensionality of the spectral dataset. During the PCA analysis, four groups of spectra were considered based on concentration ranges from 1 µM to 0.5 nM for dopamine and epinephrine . In contrast, spectra containing only deionized water (DI) and serotonin served as the blank control and negative control, respectively. Each group consisted of 10 replicate spectra, for a total of 100 spectra. PCA extracts the dominant variance patterns, and the first two principal components (PC1 and PC2) captured 99% of the total variance. This means most of the spectral information was preserved even in a low-dimensional vector space. The score plot of PC1 vs PC2 (**Figure 8**) shows a clear separation between the reference groups (DI and SE) and the targeted analyte-containing groups (DA and EP). The DA and EP spectra form distinct clusters based on their concentration levels, suggesting that the principal components effectively capture variations related to concentration. While the intra-class high clustering demonstrates strong reproducibility and minimal variation, inter-class separation is clearly discernible with negligible overlap across classes. The PCA scores were then used as inputs for linear discriminant analysis (LDA). While PCA is an unsupervised dimensionality reduction technique, LDA is a supervised classification method.[104,105] It works by projecting data into a space that maximizes the inter-class differences while minimizing the intra-class variance.

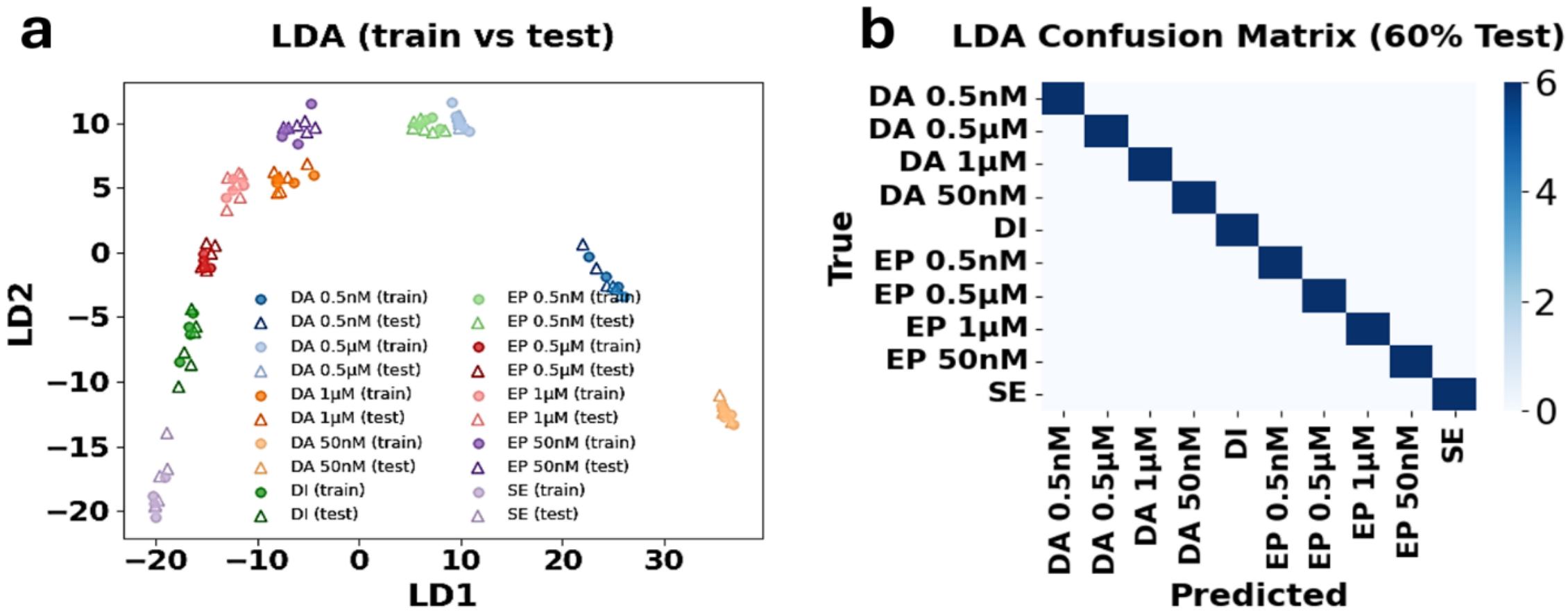

**Figure 9.** (a) LDA performed on PCA scores, showing inter-class separation, plotted with training (solid markers) and independent test sets (open markers) along the first two discriminant axes (LD1 and LD2). LDA was trained using 4-fold cross-validation within the training set while reserving 60% of the data for independent test evaluation. (b) Confusion matrix of the LDA classification on the independent test set, confirming high classification accuracy (100%) across all classes.

To enhance the model's robustness, the dataset was split into training and test subsets. During this process, 60% of the data was set aside as an unseen test set to provide an unbiased estimate of model performance. Within the training set, 4-fold cross-validation was employed to tune and validate the LDA classifier, ensuring generalization and preventing overfitting. The LDA score plot (**Figure 9a**) displays clear separation among the spectral classes, with training data (solid markers) and test data (open markers) clustering consistently by class. The corresponding confusion matrix (Figure 9b) confirms high classification accuracy (100%), with all test spectra correctly assigned to their respective groups. Notably, the model achieved 99% accuracy while using 70% of the data for testing and 30% for training (Figure S11). The effects of the number of PC's across different cross-validation methods on overall accuracy are detailed in SI (Figures S8, S9, and S10).

Our study demonstrates that combining SERS with PCA-LDA provides a robust analytical framework for evaluating spectral differences arising from catechol-mediated molecular attachment on defect-engineered $MoS_2$. The classification model creates distinct, concentration-dependent clusters with high accuracy, reflecting consistent spectral behavior. Notably, using multivariate pattern recognition instead of relying on single Raman bands reduces interference from background signals, enabling more reliable comparison of structurally similar analytes. PCA filters spectral noise by isolating the main variance related to the analyte, while LDA enhances group separation for accurate classification between closely related spectral responses. Along with the molecular selectivity of defect-engineered $MoS_2$, this framework establishes a foundation for future studies aimed at extending qualitative and quantitative SERS analysis to complex biological samples.

## 3. Conclusion

In summary, we developed an optimized soft plasma etching method using RIE to create sulfur defects with controlled defect density. We demonstrated that S-vacancies significantly modulate the structural, chemical, and optoelectronic properties of 2D $MoS_2$, establishing it as a potential SERS platform for catechol-containing neurotransmitter detection. The controllable introduction of defects was confirmed using a range of spectroscopic techniques, including Raman spectroscopy, photoluminescence, and XPS, corroborated through a detailed surface morphology study using AFM and SEM.

The defect-engineered 2D $MoS_2$ was investigated as a SERS-active platform for neurotransmitters' detection using dopamine, epinephrine, and serotonin as model analytes. The catechol-mediated dopamine and epinephrine exhibited clear SERS responses down to $10^{-10}$ M, with calibration curves exhibiting high correlation coefficients for pure samples ($R^2$ = 0.95 and 0.99, respectively) and indicating consistent concentration-dependent response. Non- catechol-containing molecule of serotonin did not exhibit any SERS spectra, demonstrating the specificity of the SERS enhancement. We proposed the mechanism of neurotransmitters' chemical attachment to defected $MoS_2$, highlighting how S vacancies influence molecular adsorption.

We demonstrated the utility of machine learning based multivariate analysis techniques, particularly PCA-LDA, for distinguishing between different neurotransmitters: dopamine and epinephrine, based on their spectral signatures. The machine learning model successfully distinguished (accuracy of 100%) spectra of dopamine and epinephrine across different concentration levels. These results demonstrate the utility of machine learning as an analytical tool for resolving highly similar spectral responses.

In conclusion, our findings establish defect engineering in $MoS_2$ as an effective strategy for activating and tuning SERS responses through catechol-mediated molecular interactions under controlled etching conditions. The insights gained from this study lay the foundation for further advances in the design of robust, cost-effective, and tunable semiconductor-based SERS platforms for selective molecular detection.

## 4. Methods

Materials. Monolayer and multilayer films of $MoS_2$ were obtained from the 2DCC-MIP.

The sample's detailed growth recipe and all characterization data is available at: https://data.2dccmip.org/pjoGS9H7rOe0. And subsequently transferred onto $SiO_2$/Si substrates using wet transfer technique described elsewhere[106,107] (SI). Dopamine hydrochloride (C8H12ClNO2), epinephrine hydrochloride (C9H14ClNO3), and serotonin hydrochloride (C10H13ClN2O) were purchased from Thermo Fisher Scientific, ensuring high purity (99%) for all experiments.

Methods. Plasma etching was performed using a PlasmaTherm ICP 770 DRIE system (gas sources: $SF_6$ and Ar). Raman and photoluminescence characterization were acquired with a WITec alpha300R confocal Raman microscope equipped with a 100× objective lens and a 532 nm laser. For $MoS_2$ characterization, Raman and photoluminescence spectra were collected using a 1 mW laser power and 1800 g/mm or 300 g/mm diffraction grating respectively. All other Raman spectra, including those for surface-enhanced Raman spectroscopy, were measured using a 2mW laser power, a 600 g/mm diffraction grating. We fitted the Raman peaks with a standard 3-parameter Lorentzian model. X-ray photoelectron spectroscopy (XPS) was conducted using a Thermo Fisher Scientific ESCALAB Xi+ system. Scanning electron microscopy (SEM) images were obtained using a Zeiss Auriga FESEM instrument with a 5 kV electron beam. Atomic force microscopy (AFM) was carried out on an Asylum MFP-3D Origin AFM in taping mode with Tap 300 Al-G tips.

2mM stock solution of dopamine, epinephrine, or serotonin, was prepared by dissolving in DI water. Solutions of lower concentrations (1mM, 1uM, 0.5um, 50nM, 0.5nM) were then prepared by serial dilution of the stock solution. Immediately after the reactor, each SERS substrate, plasma-treated MoS2, was incubated for 2 hours in a solution containing a particular concentration of a neurotransmitter (dopamine, epinephrine, or serotonin), followed by drying via a nitrogen blow in the clean room environment to prevent any contamination. Raman scattering measurements, were conducted on these samples, including the reference samples, immediately after preparation. All spectra were plotted within the 1050-1800 $cm^{-1}$ range and normalized to the Si peak intensity at 520 $cm^{-1}$.

Statistical analysis (LDA) of the data was performed using PyCharm, utilizing custom-written Python script for data processing and analysis.

**Funding**

AK, TI: the University of North Carolina System for the UNC ROI Program funding in 2023–2026; BK, TI: the US Army DEVCOM Soldier Center [Contract W911QY2220006]; NT, CC, JMR, SVR, TI: NSF DMR-2039351.

**Acknowledgements**

AK, BK, TI acknowledge the Joint School of Nanoscience and Nanoengineering, a member of the South-Eastern Nanotechnology Infrastructure Corridor (SENIC) and National Nanotechnology Coordinated Infrastructure (NNCI), supported by the NSF (Grant ECCS-1542174). $MoS_2$ samples were provided by the 2DCC-MIP RSVP program, which is supported by NSF cooperative agreement DMR-2039351.

**Author Contributions**

AK: Writing, Methodology, Investigation, Formal analysis. BK: Investigation. NT: Synthesis, Investigation. CC: Synthesis, Investigation. JMR: Supervision, Synthesis, Investigation. SVR: Supervision, Conceptualization. TI: Writing, Supervision, Conceptualization, Funding acquisition.

**Conflict of Interest**

The authors declare no conflict of interest

**References**

1 B. Chen, L. Fan, C. Li, L. Xia, K. Wang, J. Wang, D. Pang, Z. Zhu and P. Ma, *Analyst*, 2024, **149**, 4283–4294.

2 Z. Sun, Y. Gao, C. Ban, J. Meng, J. Wang, K. Wang and L. Gan, *ACS Appl. Nano Mater.*, 2023, **6**, 8635–8642.

3 Q. Zhao, H. Hilal, J. Kim, W. Park, M. N. Haddadnezhad, J. Lee, W. Park, J. W. Lee, S. Lee, I. Jung and S. Park, *J. Am. Chem. Soc.*, 2022, **144**, 13285–13293.

4 M. F. Cardinal, E. Vander Ende, R. A. Hackler, M. O. McAnally, P. C. Stair, G. C. Schatz and R. P. Van Duyne, *Chem. Soc. Rev.*, 2017, **46**, 3886–3903.

5 R. G. Freeman, K. C. Grabar, K. J. Allison, R. M. Bright, J. A. Davis, A. P. Guthrie, M. B. Hommer, M. A. Jackson, P. C. Smith, D. G. Walter and M. J. Natan, *Science (1979).*, 1995, **267**, 1629–1632.

6 S. Kaja, A. V. Mathews, V. V. K. Venuganti and A. Nag, *Langmuir*, 2023, **39**, 5591–5601.

7 P. Dai, H. Li, X. Huang, N. Wang and L. Zhu, *Nanomaterials 2021, Vol. 11, Page 2770*, 2021, **11**, 2770.

8 T. J. Wang, N. R. Barveen, Z. Y. Liu, C. H. Chen and M. H. Chou, *ACS Appl. Mater. Interfaces*, 2021, **13**, 34910–34922.

9 Y. Xu, Q. Dong, S. Cong and Z. Zhao, *Analysis & Sensing*, 2024, **4**, e202300067.

10 Y. Chen, H. Wang, J. Zhou, D. Lin, L. Zhang, Z. Xing, Q. Zhang and L. Xia, *Spectrochim. Acta A Mol. Biomol. Spectrosc.*, 2024, **318**, 124487.

11 S. Al Abdullah, S. Ghadami, M. Arifur Rahman Khan, F. Ebrahimi, K. Nowlin, T. Ignatova, K. Dellinger, S. Al Abdullah, S. Ghadami, F. Ebrahimi, K. Dellinger, M. A. R Khan, K. Nowlin and T. Ignatova, *Advanced Sensor Research*, 2025, **4**, e00034.

12 F. Munshe and Md. A. R. Khan, *Nanoscience & Nanotechnology-Asia*, DOI:10.2174/2210681212666220618164341.

13 F. M. Zablon, M. Arifur, R. Khan, T. Ignatova, K. Dellinger and S. Aravamudhan, DOI:10.1002/admi.202500528.

14 L. Yang, Y. Peng, Y. Yang, J. Liu, H. Huang, B. Yu, J. Zhao, Y. Lu, Z. Huang, Z. Li and J. R. Lombardi, *Advanced Science*, 2019, **6**, 1900310.

15 H. Wei, Z. Peng, C. Yang, Y. Tian, L. Sun, G. Wang and M. Liu, *Nanomaterials*, 2021, **11**, 2026.

16 W. Zou, Z. Wan, X. Yu, Z. Liu, P. Yuan and X. Zhang, *J. Hazard. Mater.*, 2021, **419**, 126499.

17 Z. Zheng, S. Cong, W. Gong, J. Xuan, G. Li, W. Lu, F. Geng and Z. Zhao, *Nature Communications 2017 8:1*, 2017, **8**, 1–10.

18 Y. Chen, X. Wang, G. Wu, Z. Wang, H. Fang, T. Lin, S. Sun, H. Shen, W. Hu, J. Wang, J. Sun, X. Meng and J. Chu, *Small*, 2018, **14**, 1703293.

19 H. He, F. Yang, Z. Kuang, M. Yang, Y. Yu, A. Wang, J. Mao, R. Shu, Y. Luo, Z. Xie, M. Tian, J. Zheng, M. Wang, Y. Huang, Y. Su, L. Du, Q. Zhang, D. Li, Z. C. Wang, L. Li and J. Zhu, *ACS Appl. Nano Mater.*, DOI:10.1021/acsanm.4c04529.

20 X. Ling, L. Xie, Y. Fang, H. Xu, H. Zhang, J. Kong, M. S. Dresselhaus, J. Zhang and Z. Liu, *Nano Lett.*, 2010, **10**, 553–561.

21 P. Wang, M. Xia, O. Liang, K. Sun, A. F. Cipriano, T. Schroeder, H. Liu and Y. H. Xie, *Anal. Chem.*, 2015, **87**, 10255–10261.

22 Z. Yin, K. Xu, S. Jiang, D. Luo, R. Chen, C. Xu, P. Shum and Y. J. Liu, *Materials Today Physics*, 2021, **18**, 100378.

23 P. Karthick Kannan, P. Shankar, C. Blackman and C. H. Chung, *Advanced Materials*, 2019, **31**, 1803432.

24 J. Kim, Y. Jang, N. J. Kim, H. Kim, G. C. Yi, Y. Shin, M. H. Kim and S. Yoon, *Front. Chem.*, DOI:10.3389/fchem.2019.00582.

25 C. Muehlethaler, C. R. Considine, V. Menon, W. C. Lin, Y. H. Lee and J. R. Lombardi, *ACS Photonics*, 2016, **3**, 1164–1169.

26 H. Kitadai, X. Wang, N. Mao, S. Huang and X. Ling, *Journal of Physical Chemistry Letters*, 2019, **10**, 3043–3050.

27 L. Tao, K. Chen, Z. Chen, C. Cong, C. Qiu, J. Chen, X. Wang, H. Chen, T. Yu, W. Xie, S. Deng and J. Bin Xu, *J. Am. Chem. Soc.*, 2018, **140**, 8696–8704.

28 Y. Quan, J. Yao, Y. Sun, X. Qu, R. Su, M. Hu, L. Chen, Y. Liu, M. Gao and J. Yang, *Sens. Actuators B Chem.*, 2021, **327**, 128903.

29 Q. Cai, L. H. Li, Y. Yu, Y. Liu, S. Huang, Y. Chen, K. Watanabe and T. Taniguchi, *Physical Chemistry Chemical Physics*, 2015, **17**, 7761–7766.

30 X. Ling, W. Fang, Y. H. Lee, P. T. Araujo, X. Zhang, J. F. Rodriguez-Nieva, Y. Lin, J. Zhang, J. Kong and M. S. Dresselhaus, *Nano Lett.*, 2014, **14**, 3033–3040.

31 Z. He, T. Rong, Y. Li, J. Ma, Q. Li, F. Wu, Y. Wang and F. Wang, *ACS Nano*, 2022, **16**, 4072–4083.

32 T. Ignatova, S. Pourianejad, X. Li, K. Schmidt, F. Aryeetey, S. Aravamudhan and S. V. Rotkin, *ACS Nano*, 2022, **16**, 2598–2607.

33 J. R. Lombardi and R. L. Birke, *Journal of Physical Chemistry C*, 2014, **118**, 11120–11130.

34 S. H. Amsterdam, T. K. Stanev, Q. Zhou, A. J. T. Lou, H. Bergeron, P. Darancet, M. C. Hersam, N. P. Stern and T. J. Marks, *ACS Nano*, 2019, **13**, 4183–4190.

35 X. Ling, L. G. Moura, M. A. Pimenta and J. Zhang, *Journal of Physical Chemistry C*, 2012, **116**, 25112–25118.

36 D. Glass, E. Cortés, S. Ben-Jaber, T. Brick, W. J. Peveler, C. S. Blackman, C. R. Howle, R. Quesada-Cabrera, I. P. Parkin and S. A. Maier, *Advanced Science*, 2019, **6**, 1901841.

37 X. Wang, W. Shi, S. Wang, H. Zhao, J. Lin, Z. Yang, M. Chen and L. Guo, *J. Am. Chem. Soc.*, 2019, **141**, 5856–5862.

38 S. Cong, Y. Yuan, Z. Chen, J. Hou, M. Yang, Y. Su, Y. Zhang, L. Li, Q. Li, F. Geng and Z. Zhao, *Nature Communications 2015 6:1*, 2015, **6**, 1–7.

39 L. Yang, D. Yin, Y. Shen, M. Yang, X. Li, X. Han, X. Jiang and B. Zhao, *Physical Chemistry Chemical Physics*, 2017, **19**, 22302–22308.

40 L. Yang, Y. Peng, Y. Yang, J. Liu, Z. Li, Y. Ma, Z. Zhang, Y. Wei, S. Li, Z. Huang and N. V. Long, *ACS Appl. Nano Mater.*, 2018, **1**, 4516–4527.

41 X. Hou, X. Luo, X. Fan, Z. Peng and T. Qiu, *Physical Chemistry Chemical Physics*, 2019, **21**, 2611–2618.

42 X. Jiang, Q. Sang, M. Yang, J. Du, W. Wang, L. Yang, X. Han and B. Zhao, *Physical Chemistry Chemical Physics*, 2019, **21**, 12850–12858.

43 L. Yang, Y. Peng, Y. Yang, J. Liu, H. Huang, B. Yu, J. Zhao, Y. Lu, Z. Huang, Z. Li and J. R. Lombardi, *Advanced Science*, 2019, **6**, 1900310.

44 J. Ye, R. Arul, M. K. Nieuwoudt, J. Dong, T. Zhang, L. Dai, N. C. Greenham, A. Rao, R. L. Z. Hoye, W. Gao and M. C. Simpson, *Journal of Physical Chemistry Letters*, 2023, 4607–4616.

45 X. Y. Zhang, D. Han, Z. Pang, Y. Sun, Y. Wang, Y. Zhang, J. Yang and L. Chen, *Journal of Physical Chemistry C*, 2018, **122**, 5599–5605.

46 M. Li, X. Fan, Y. Gao and T. Qiu, *Journal of Physical Chemistry Letters*, 2019, **10**, 4038–4044.

47 Y. Wang, J. Liu, Y. Ozaki, Z. Xu and B. Zhao, *Angewandte Chemie*, 2019, **131**, 8256–8260.

48 L. Yang, Q. Sang, J. Du, M. Yang, X. Li, Y. Shen, X. Han, X. Jiang and B. Zhao, *Physical Chemistry Chemical Physics*, 2018, **20**, 15149–15157.

49 E. Feng, T. Zheng, X. He, J. Chen and Y. Tian, *Sci. Adv.*, DOI:10.1126/SCIADV.AAU3494/SUPPL_FILE/AAU3494_SM.PDF.

50 S. Almohammed, F. Zhang, B. J. Rodriguez and J. H. Rice, *Scientific Reports 2018 8:1*, 2018, **8**, 1–10.

51 X. Zheng, F. Ren, S. Zhang, X. Zhang, H. Wu, X. Zhang, Z. Xing, W. Qin, Y. Liu and C. Jiang, *ACS Appl. Mater. Interfaces*, 2017, **9**, 14534–14544.

52 H. Kitadai, Q. Tan, L. Ping and X. Ling, *NPJ 2D Mater. Appl.*, DOI:10.1038/s41699-024-00446-z.

53 Y. C. Lin, R. Torsi, R. Younas, C. L. Hinkle, A. F. Rigosi, H. M. Hill, K. Zhang, S. Huang, C. E. Shuck, C. Chen, Y. H. Lin, D. Maldonado-Lopez, J. L. Mendoza-Cortes, J. Ferrier, S. Kar, N. Nayir, S. Rajabpour, A. C. T. Van Duin, X. Liu, D. Jariwala, J. Jiang, J. Shi, W. Mortelmans, R. Jaramillo, J. M. J. Lopes, R. Engel-Herbert, A. Trofe, T. Ignatova, S. H. Lee, Z. Mao, L. Damian, Y. Wang, M. A. Steves, K. L. Knappenberger, Z. Wang, S. Law, G. Bepete, D. Zhou, J. X. Lin, M. S. Scheurer, J. Li, P. Wang, G. Yu, S. Wu, D. Akinwande, J. M. Redwing, M. Terrones and J. A. Robinson, *ACS Nano*, 2023, **17**, 9694–9747.

54 X. Tang, Q. Hao, X. Hou, L. Lan, M. Li, L. Yao, X. Zhao, Z. Ni, X. Fan and T. Qiu, *John Wiley and Sons Inc*, 2024, preprint, DOI: 10.1002/adma.202312348.

55 M. Rajput, S. K. Mallik, S. Chatterjee, A. Shukla, S. Hwang, S. Sahoo, G. V. P. Kumar and A. Rahman, *Commun. Mater.*, DOI:10.1038/s43246-024-00632-y.

56 F. Aryeetey, S. Pourianejad, O. Ayanbajo, K. Nowlin, T. Ignatova and S. Aravamudhan, *RSC Adv.*, 2021, **11**, 20893–20898.

57 S. Mouri, Y. Miyauchi and K. Matsuda, *Nano Lett.*, 2013, **13**, 5944–5948.

58 R. Thayil, S. R. Parne and C. V. Ramana, *Small*, 2025, **21**, 2412467.

59 R. Jha, P. Mishra and S. Kumar, *Elsevier Ltd*, 2024, preprint, DOI: 10.1016/j.bios.2024.116232.

60 R. Ge, X. Wu, M. Kim, J. Shi, S. Sonde, L. Tao, Y. Zhang, J. C. Lee and D. Akinwande, *Nano Lett.*, 2018, **18**, 434–441.

61 K. F. Mak and J. Shan, *Nat. Photonics*, 2016, **10**, 216–226.

62 M. F. Hossen, S. Shendokar, Md. A. R. Khan and S. Aravamudhan, *Nano Select*, DOI:10.1002/nano.202400103.

63 C. Tsai, H. Li, S. Park, J. Park, H. S. Han, J. K. Nørskov, X. Zheng and F. Abild-Pedersen, *Nature Communications 2017 8:1*, 2017, **8**, 1–8.

64 Q. Zhang, A. T. S. Wee, Q. Liang, X. Zhao and M. Liu, *ACS Nano*, 2021, **15**, 2165–2181.

65 P. Zuo, L. Jiang, X. Li, P. Ran, B. Li, A. Song, M. Tian, T. Ma, B. Guo, L. Qu and Y. Lu, *Nanoscale*, 2019, **11**, 485–494.

66 R. Rani, A. Yoshimura, S. Das, M. R. Sahoo, A. Kundu, K. K. Sahu, V. Meunier, S. K. Nayak, N. Koratkar and K. S. Hazra, *ACS Nano*, 2020, **14**, 6258–6268.

67 S. Bertolazzi, S. Bonacchi, G. Nan, A. Pershin, D. Beljonne and P. Samorì, *Advanced Materials*, 2017, **29**, 1606760.

68 C. Brownlee, *ACS Nano*, 2017, **11**, 3425–3428.

69 Y. T. Lei, D. W. Li, T. C. Zhang, X. Huang, L. Liu and Y. F. Lu, *J. Mater. Chem. C Mater.*, 2017, **5**, 8883–8892.

70 L. P. de Vries, M. P. van de Weijer and M. Bartels, *Elsevier Ltd*, 2022, preprint, DOI: 10.1016/j.neubiorev.2022.104733.

71 L. Speranza, U. Di Porzio, D. Viggiano, A. De Donato and F. Volpicelli, DOI:10.3390/cells.

72 M. Kim, Y. S. Choi and D. H. Jeong, *RSC Adv.*, 2024, **14**, 14214–14220.

73 M. Rajput, S. K. Mallik, S. Chatterjee, A. Shukla, S. Hwang, S. Sahoo, G. V. P. Kumar and A. Rahman, *Commun. Mater.*, DOI:10.1038/s43246-024-00632-y.

74 T. Z. Lin, B. T. Kang, M. H. Jeon, C. Huffman, J. H. Jeon, S. J. Lee, W. Han, J. Y. Lee, S. H. Lee, G. Y. Yeom and K. N. Kim, *ACS Appl. Mater. Interfaces*, 2015, **7**, 15892–15897.

75 S. Xiao, P. Xiao, X. Zhang, D. Yan, X. Gu, F. Qin, Z. Ni, Z. J. Han and K. K. Ostrikov, *Sci. Rep.*, 2016, **6**, 1–8.

76 B. J. Chou, Y. Y. Chung, W. S. Yun, C. F. Hsu, M. Y. Li, S. K. Su, S. L. Liew, V. D. H. Hou, C. W. Chen, C. C. Kei, Y. Y. Shen, W. H. Chang, T. Y. Lee, C. C. Cheng, I. P. Radu and C. H. Chien, *Nanotechnology*, 2024, **35**, 125204.

77 H. Kii, N. Nagamura and R. Nouchi, *Nano Express*, 2025, **6**, 025008.

78 H. Zhu, X. Qin, L. Cheng, A. Azcatl, J. Kim and R. M. Wallace, *ACS Appl. Mater. Interfaces*, 2016, **8**, 19119–19126.

79 C. Lee, H. Yan, L. E. Brus, T. F. Heinz, J. Hone and S. Ryu, *ACS Nano*, 2010, **4**, 2695–2700.

80 R. Dhall, M. R. Neupane, D. Wickramaratne, M. Mecklenburg, Z. Li, C. Moore, R. K. Lake and S. Cronin, *Advanced Materials*, 2015, **27**, 1573–1578.

81 B. Chakraborty, H. S. S. R. Matte, A. K. Sood and C. N. R. Rao, *Journal of Raman Spectroscopy*, 2013, **44**, 92–96.

82 S. Mignuzzi, A. J. Pollard, N. Bonini, B. Brennan, I. S. Gilmore, M. A. Pimenta, D. Richards and D. Roy, *Phys. Rev. B Condens. Matter Mater. Phys.*, 2015, **91**, 195411.

83 T. Grünleitner, A. Henning, M. Bissolo, M. Zengerle, L. Gregoratti, M. Amati, P. Zeller, J. Eichhorn, A. V. Stier, A. W. Holleitner, J. J. Finley and I. D. Sharp, *ACS Nano*, 2022, **16**, 20364–20375.

84 K. F. Mak, K. He, C. Lee, G. H. Lee, J. Hone, T. F. Heinz and J. Shan, *Nat. Mater.*, 2013, **12**, 207–211.

85 N. Scheuschner, O. Ochedowski, A. M. Kaulitz, R. Gillen, M. Schleberger and J. Maultzsch, *Phys. Rev. B Condens. Matter Mater. Phys.*, 2014, **89**, 125406.

86 L. A. H. Jones, Z. Xing, J. E. N. Swallow, H. Shiel, T. J. Featherstone, M. J. Smiles, N. Fleck, P. K. Thakur, T. L. Lee, L. J. Hardwick, D. O. Scanlon, A. Regoutz, T. D. Veal and V. R. Dhanak, *Journal of Physical Chemistry C*, 2022, **126**, 21022–21033.

87 H. Mohammad-Shiri, M. Ghaemi, S. Riahi and A. Akbari-Sehat, *Computational and Electrochemical Studies on the Redox Reaction of Dopamine in Aqueous Solution*, 2011, vol. 6.

88 J. R. Lombardi, R. L. Birke, T. Lu and J. Xu, *J. Chem. Phys.*, 1986, **84**, 4174–4180.

89 J. R. Lombardi and R. L. Birke, *Journal of Physical Chemistry C*, 2008, **112**, 5605–5617.

90 J. R. Lombardi and R. L. Birke, *Journal of Chemical Physics*, DOI:10.1063/1.2748386/296059.

91 Z. Jamshidi, S. Ashtari-Jafari, A. Smirnov and E. V. Solovyeva, *Journal of Physical Chemistry C*, 2021, **125**, 17202–17211.

92 V. Hugo, C. Castro, C. Lucía, L. Valenzuela, J. Carlos Salazar Sánchez, K. Pardo Peña, S. J. López Pérez, J. O. Ibarra and A. Morales Villagrán, *Send Orders for Reprints to reprints@benthamscience.ae An Update of the Classical and Novel Methods Used for Measuring Fast Neurotransmitters During Normal and Brain Altered Function*, 2014, vol. 12.

93 P. G. Spizzirri, J.-H. Fang, S. Rubanov, E. Gauja and S. Prawer, *Nano-Raman spectroscopy of silicon surfaces*, .

94 K. W. Kho, U. S. Dinish, A. Kumar and M. Olivo, *ACS Nano*, 2012, **6**, 4892–4902.

95 N. Valley, N. Greeneltch, R. P. Van Duyne and G. C. Schatz, *Journal of Physical Chemistry Letters*, 2013, **4**, 2599–2604.

96 Pamela. Córdova, .

97 V. T. Vo, V. D. Phung and S. W. Lee, *Surfaces and Interfaces*, 2021, **25**, 101181.

98 Y. Gwon, J. H. Kim and S. W. Lee, *ACS Sens.*, 2024, **9**, 870–882.

99 J. Chi, Y. Ma, F. L. Weng, H. Thiessen-Philbrook, C. R. Parikh and H. Du, *J. Biophotonics*, 2017, **10**, 1743–1755.

100 M. Marro, A. Taubes, P. Villoslada and D. Petrov, *https://doi.org/10.1117/12.921358*, 2012, **8427**, 228–234.

101 B. Sharma, P. Bugga, L. R. Madison, A. I. Henry, M. G. Blaber, N. G. Greeneltch, N. Chiang, M. Mrksich, G. C. Schatz and R. P. Van Duyne, *J. Am. Chem. Soc.*, 2016, **138**, 13952–13959.

102 W. R. Premasiri, Y. Gebregziabher and L. D. Ziegler, *http://dx.doi.org/10.1366/10-06173*, 2011, **65**, 493–499.

103 A. K. Boardman, W. S. Wong, W. R. Premasiri, L. D. Ziegler, J. C. Lee, M. Miljkovic, C. M. Klapperich, A. Sharon and A. F. Sauer-Budge, *Anal. Chem.*, 2016, **88**, 8026–8035.

104 DOI:10.1007/978-0-387-78189-1_.

105 A. Tharwat, T. Gaber, A. Ibrahim and A. E. Hassanien, *AI Communications*, 2017, **30**, 169–190.

106 M. A. R. Khan, S. Kalkar, B. Khader, O. O. Ayodele, A. Prokofjevs and T. Ignatova, *Nanoscale*, 2025, **17**, 19870–19881.

107 A. ’Sheets, S. ’Kumari, ‘Chen Chen’, A. ’Graves, J. M. ’Redwing and T. ’Ignatova, in *91st Annual Meeting of the Southeastern Section of the APS*, Bulletin of the American Physical Society, Charlotte, NC, USA.

Supporting Information

# Defect Engineered 2D $MoS_2$ Materials for ML-enabled Neurotransmitter SERS Detection

*Md Arifur R. Khan, Besan Khader, Nicholas Trainor, Chen Chen, Joan M Redwing, Slava V Rotkin, and Tetyana Ignatova**

- Transfer protocol of $MoS_2$ from sapphire to the target substrate.
- Details on characterization plasma treated $MoS_2$. Defect density calculations.
- SERS measurements.
- Machine learning based multivariate analysis.
- Python code for the LDA model.

## 1. RANSFER PROTOCOL OF $MOS_2$ FROM SAPPHIRE TO THE TARGET SUBSTRATE

The $MoS_2$ samples used for this study were grown at The Pennsylvania State University Two-Dimensional Crystal Consortium – Materials Innovation Platform (2DCC-MIP) via metal-organic chemical vapor deposition (MOCVD) on a sapphire substrate. As-received $MoS_2$ was transferred onto a target substrate ($SiO_2$/Si) following our previous method[1], using a polymethyl methacrylate (PMMA A6) support layer. The PMMA film was applied through a two-step spin coating process. First, a PMMA A6 solution was spin-coated at 400 rpm for 15 seconds to ensure the polymer spread uniformly across the surface. Then, the spinning speed was increased to 500 rpm for 45 seconds. These parameters were chosen to enhance uniformity and control thickness. After coating, the sample was heated on a hot plate at 100 °C for 30 minutes to partially cure the PMMA.

A second layer of PMMA was applied under the same conditions to enhance the film's mechanical stability. It was then allowed to dry naturally for 24 hours at room temperature to improve adhesion between the $MoS_2$ layer and the PMMA. For delamination, thermal tape was attached to the PMMA/$MoS_2$/sapphire structure, which was then sonicated for 80 minutes in DI water at 70 °C. This step enabled the separation of the PMMA/$MoS_2$ film from the sapphire substrate. The detached film, still supported by thermal tape, was carefully placed onto the target $SiO_2$/Si substrate. A heating step at 150 °C on a hot plate was performed for 3-5 minutes to promote strong adhesion of the $MoS_2$ to the substrate. This post-transfer heating process allows controlled removal of the thermal tape. Finally, the PMMA support layer was removed by immersing the sample in an acetone bath for 30 minutes with sonication, resulting in a clean transfer of $MoS_2$ onto the $SiO_2$/Si substrate.

This PMMA-assisted transfer method provided mechanical stability throughout the process and enabled high-quality transfer suitable for device fabrication.

## 2. DETAILS ON CHARACTERIZATION PLASMA TREATED $MOS_2$. DEFECT DENSITY CALCULATIONS.

**Plasma-Induced Quenching of Raman and PL Signals**

Raman mapping was performed over a 15 µm × 15 µm area both before and after plasma etching, and the data were fitted with a Gaussian function. The plasma etching process caused a significant decrease in the intensity of the $E_{2g}$ and $A_{1g}$ Raman modes, with approximately a 50% reduction observed, as shown in Figures S1a and S1b, respectively. Figure S1c shows a representative Raman spectrum highlighting this intensity reduction under plasma-modified conditions. Additionally, the intensity of photoluminescence (PL) was noticeably reduced after plasma treatment, as demonstrated in Figure S1d.

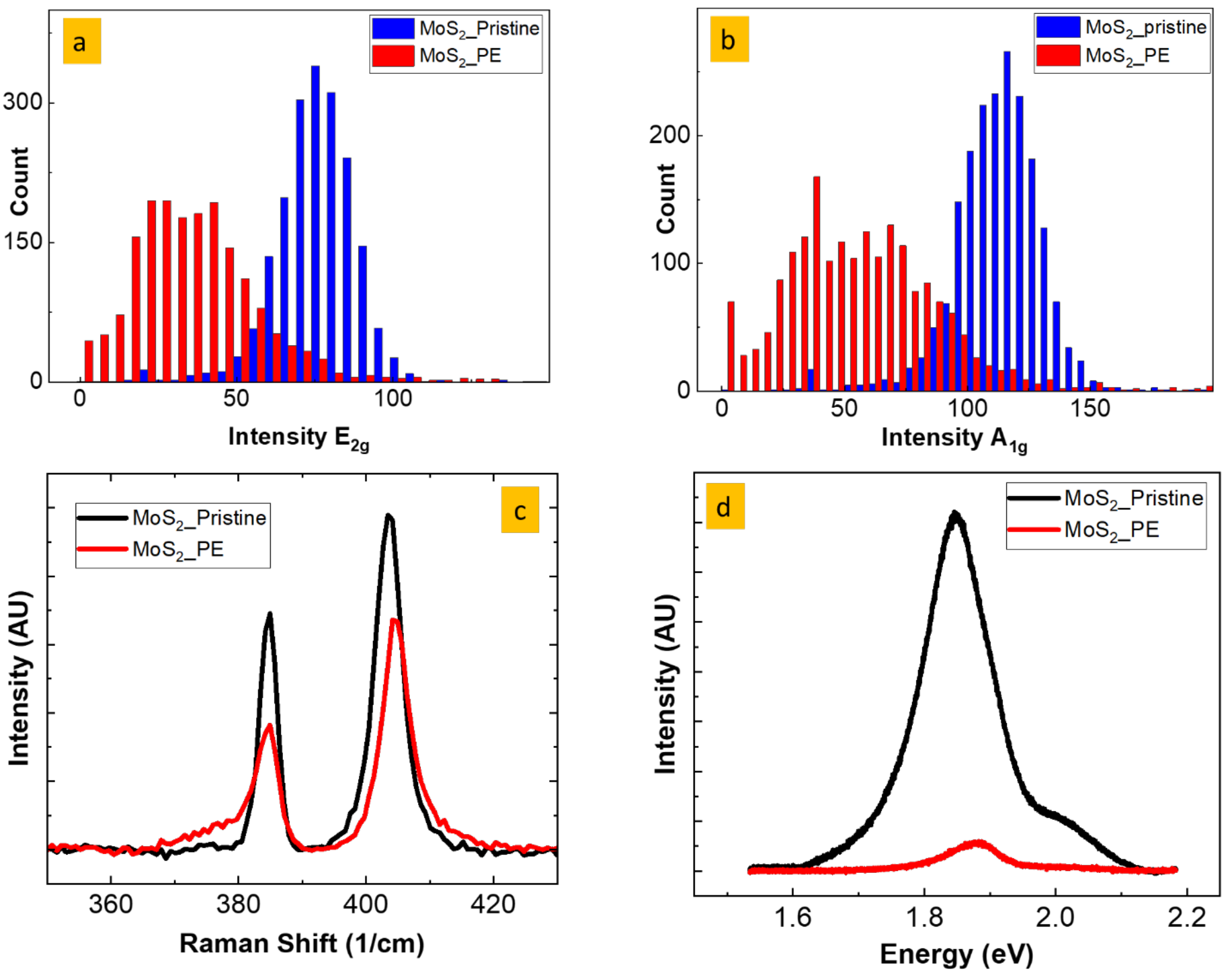

**Figure S1:** Intensity distribution of Micro Raman Mapping analysis of a 15µm x 15µm area before and after plasma etching for $E_{2g}$ (a) and $A_{1g}$ (b). Reduced intensity of Raman shift (c) and photoluminescence (d), after plasma etching (PE).

### SEM Characterization

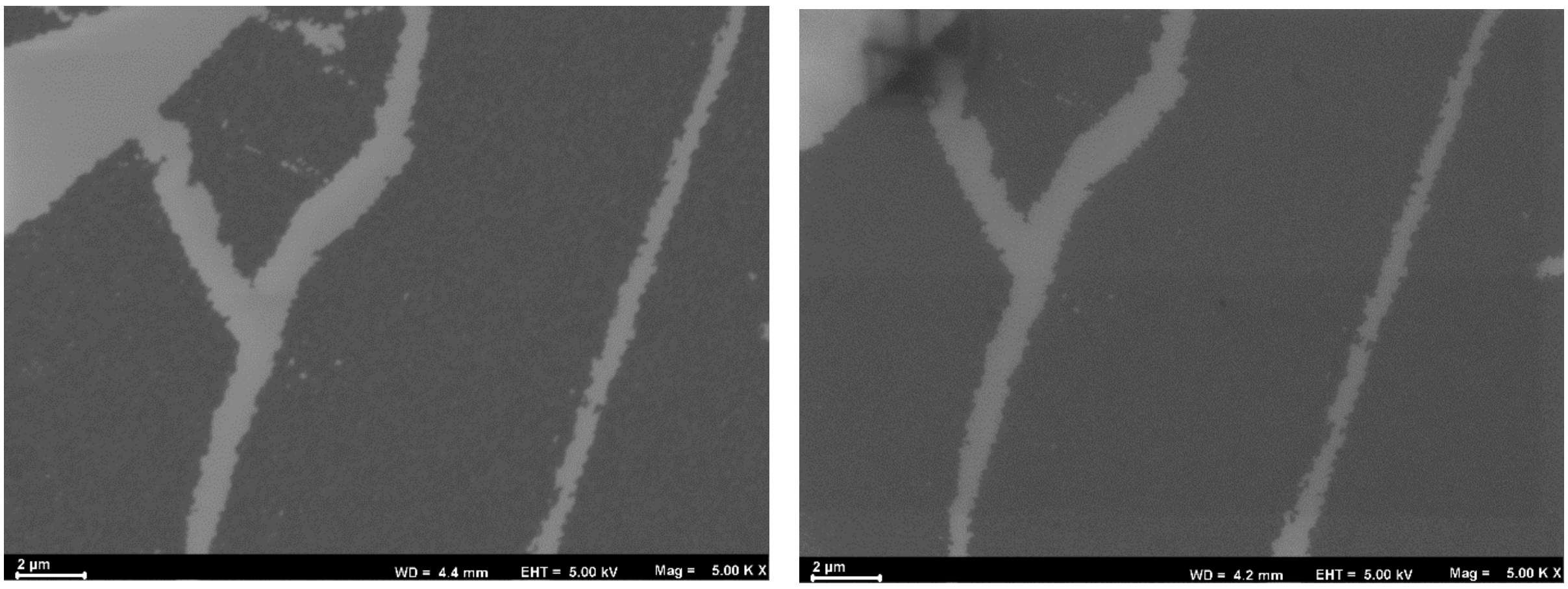

**Figure S2:** SEM Images of the same region of multilayer $MoS_2$ before (left) and after(right) soft plasma treatment.

### XPS analysis of $MoS_2$ before and after plasma etching

To investigate the chemical states and local bonding environments, along with the defect density caused by plasma treatment, X-ray Photoelectron Spectroscopy was performed on the same $MoS_2$ samples (pristine and plasma-treated) that were analyzed by SEM and AFM and discussed in the main paper. The core level Mo 3d spectra of pristine $MoS_2$ show the spin-orbit split doublet from Mo-S bonds at 230.03 eV ($Mo^{+4}$ $3d_{5/2}$) and 233.15 eV ($Mo^{+4}$ $3d_{3/2}$), as shown in Figure S3a, consistent with the literature[2–6]. Additionally, low-intensity peaks corresponding to $Mo^{+6}$, likely from the Mo precursor, appeared at a higher binding energy, while the peak at 227.1 eV is attributed to the S 2s core level[3,4,7–11]. Plasma etching induced a ~0.6 eV shift to lower binding energies in the Mo 3d doublets of the $MoS_2$ XPS spectra, indicating changes in the chemical environment of Mo and S atoms (Figure S3b). This shift has been reported as an indicator of lower oxidation states of Mo atoms (< +4) [5,7], as the local bonding environment of the Mo atoms changes when sulfur atoms are removed. Under-coordinated Mo atoms likely have a lower effective charge, possibly approaching $Mo^{2+}$ or $Mo^0$. Additional doublet peaks emerged at higher binding energies (232.14 eV and 235.49 eV) and can be attributed to these under-coordinated Mo atoms (Figure S3b). Furthermore, the main peaks of $MoS_2$ become significantly broader in the plasma-treated sample, probably due to an increased coulombic disorder in the sulfur-deficient lattice ($MoS_{2-x}$) [7,11] and a binding energy variation across different atomic sites. Notably, the $Mo^{+6}$ peak became more pronounced, likely due to oxidation resulting from atmospheric exposure during post-plasma treatment handling and XPS sample preparation. Additionally, the area ratio of the S 2s peak relative to the Mo 3d peak decreases in the defected $MoS_2$, from 0.42 to 0.32. This reduction is consistent with the selective removal of top-plane sulfur atoms which is further corroborated by defect density calculation. Figure S3c shows the survey spectra of before and after plasma etching of $MoS_2$.

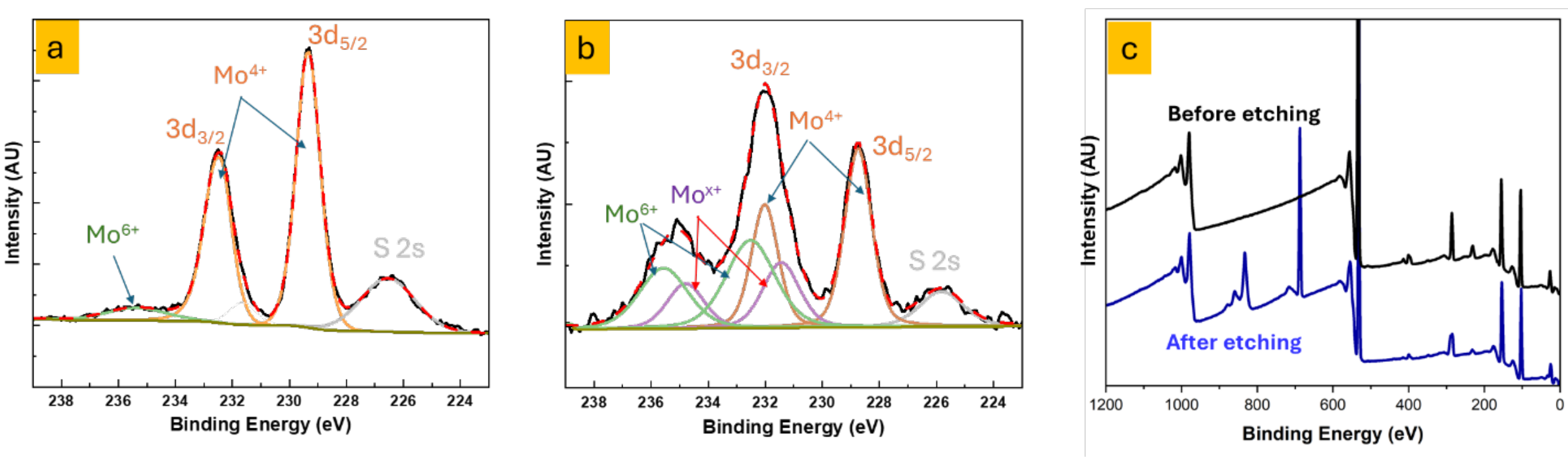

**Figure S3:** XPS spectra of Mo 3d region for $MoS_2$ before (a) and after (b) plasma etching. XPS survey spectra (c) of $MoS_2$ before (black) and after (blue) plasma etching.

Although $Mo^{6+}$ species were observed in XPS analysis, the SERS enhancement is attributed primarily to sulfur vacancies rather than surface oxides or nonspecific interaction. To minimize vacancy passivation by oxidation, analyte incubation was performed immediately after plasma treatment. Furthermore, control experiments with serotonin, which lacks catechol functionality, produced a negligible SERS response on plasma-treated $MoS_2$, supporting the conclusion that the enhancement originates from catechol coordination at undercoordinated Mo sites associated with sulfur vacancies.

Unlike conventional plasmonic substrates that primarily rely on electromagnetic enhancement, the defect-engineered $MoS_2$ platform operates mainly through charge-transfer-mediated chemical enhancement at sulfur vacancy sites, enabling selective molecular interactions. The MOCVD-grown large-area $MoS_2$ films and controlled plasma treatment provided reproducible defect formation and reasonably uniform SERS responses across multiple measurement locations, demonstrating acceptable substrate homogeneity and fabrication repeatability.

The C 1s XPS spectrum in Figure S4a of transferred and untreated $MoS_2$ confirmed the absence of polymer residuals. However, the same $MoS_2$ sample after plasma treatment showed a multicomponent C 1s XPS spectrum (Figure S4b). We attributed the appearance of new peaks to $CF_x$ species[12], originating from fluorocarbon-related surface interactions during the $SF_6$-based etching process. To confirm this statement, we performed plasma treatment on the cleanroom-grade $SiO_2$/Si wafer (Figure S4c) and obtained similar results.

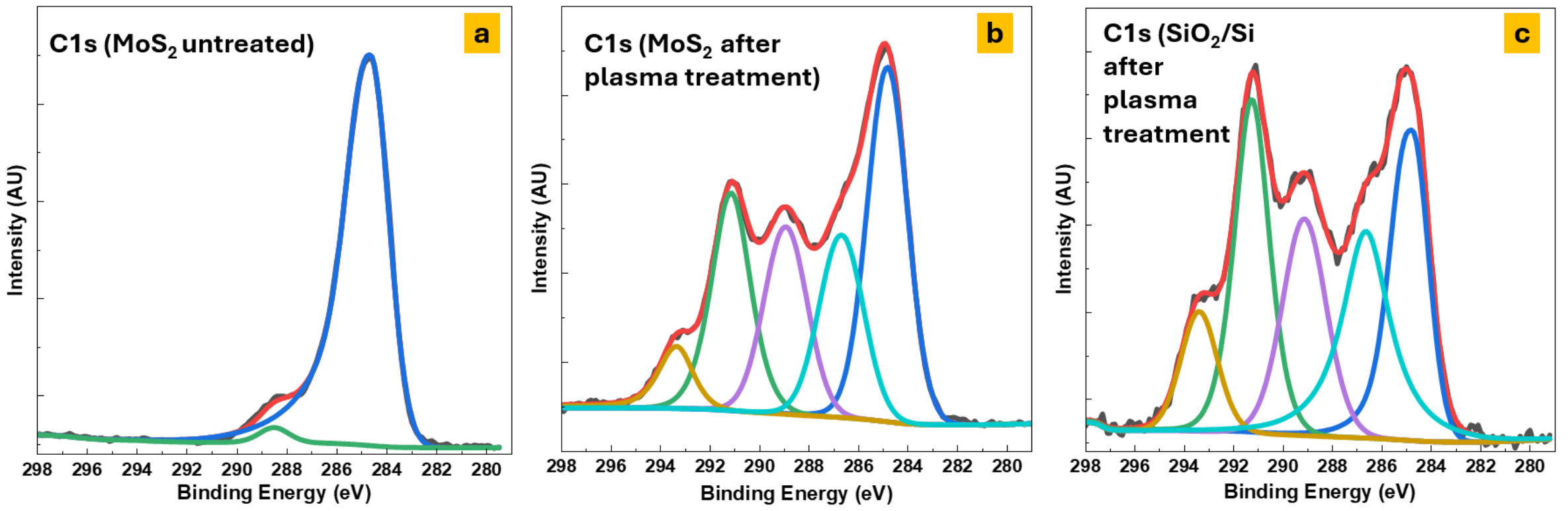

**Figure S4:** C 1s XPS spectra and peak deconvolution of (a) untreated $MoS_2$, (b) plasma-treated $MoS_2$, and (c) plasma-treated $SiO_2$/Si substrate.

Additionally, the Raman spectrum of the $MoS_2$ sample (Figure S5) showed no detectable peaks in the spectral region 1100–1800 cm−1, confirming the absence of significant residual polymer or interfering molecular contaminants prior to analyte incubation.

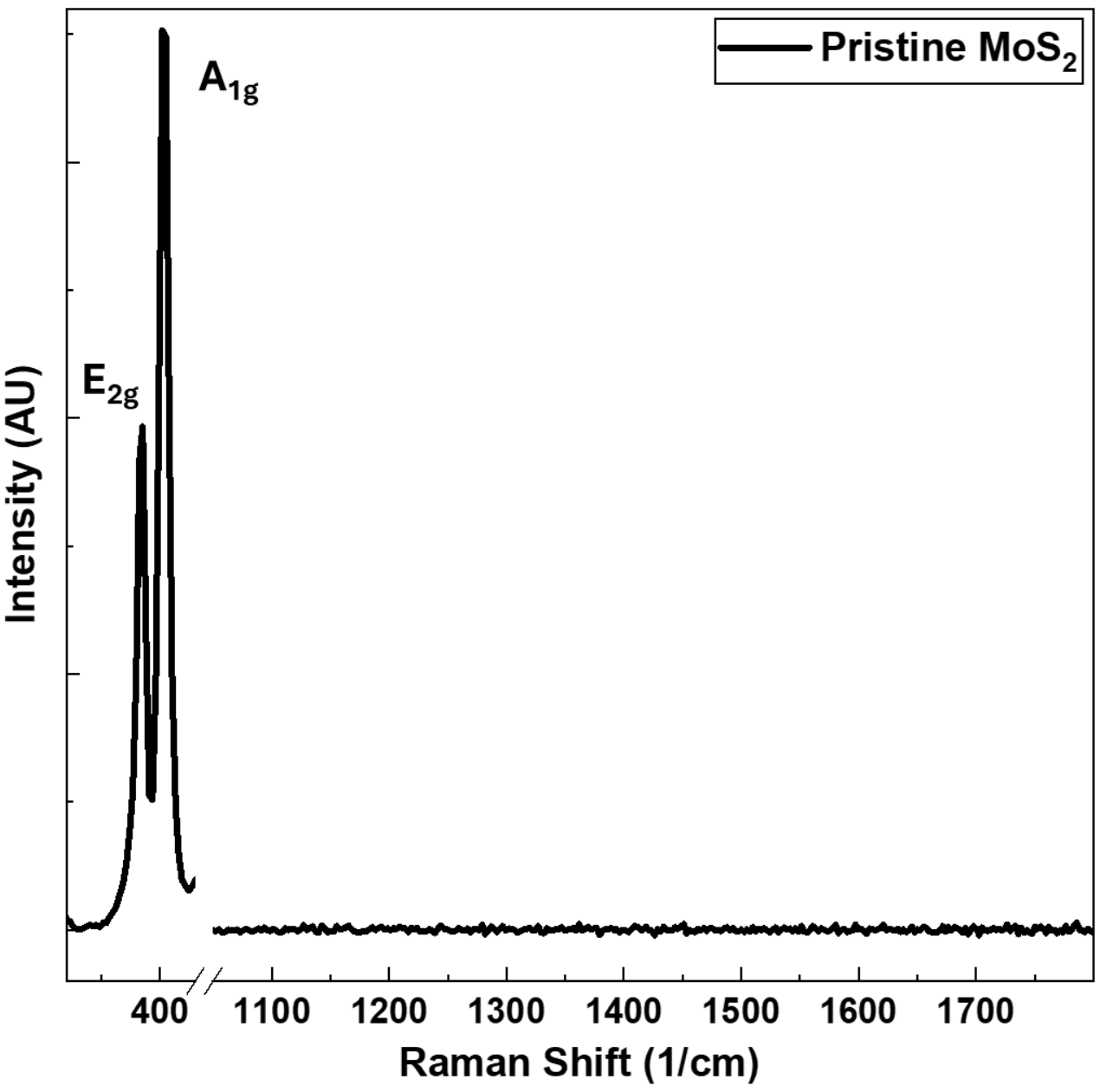

**Figure S5:** Raman spectrum of $MoS_2$ transferred onto the $SiO_2$/Si substrate, showing the characteristic $E_{2g}$ and $A_{1g}$ vibrational modes of $MoS_2$. No detectable Raman features were observed in the SERS-active region (1100–1800 $cm^{-1}$), confirming the absence of significant polymer residue or interfering molecular species prior to plasma treatment and analyte incubation.

**Defect density calculation**

The S defect density of the $MoS_2$ film was calculated for both pristine and plasma-treated samples using core-level spectra of $S_{2s}$ and $Mo_{3d}$. The stoichiometric calculation of $MoS_2$ was performed using the following equation[13,14].

$$\frac{S\,(at.\%)}{Mo(at.\%)} = \frac{I_{S2s}/\sigma_{S2s}}{I_{Mo3d5/2}/\sigma_{Mo3d5/2}}$$

In this equation, $S\,(at.\%)$ and $Mo(at.\%)$ represent the atomic percent of S and Mo, respectively. $I_{S2s}$ and $I_{Mo3d5/2}$ are the integrated area of $S_{2s}$ and $Mo_{3d5/2}$ peaks, while $\sigma_{S2s}$ and $\sigma_{Mo3d5/2}$ denote the photoionization cross sections, valued at 1.9066 and 7.4630, respectively, according to the Scofield model[15]. Additionally, considering the sulfur-sulfur distance 3.162 Å in the ideal $MoS_2$ lattice, the sulfur atom density was estimated to be $2.3 \times 10^{15}$

cm$^{-2}$ [16]. After careful calculation, we found that these considerations lead us to an S/Mo ratio of approximately 2, with a negligible S defect density. According to the previous report on realistic experimental samples, this value is consistent with the findings of a previous study[17]. In the plasma-treated $MoS_2$, the S/Mo ratio was calculated to be 1.51, with a Sulfur defect density of $5.6 \times 10^{14}$ cm$^{-2}$. These results illustrate that the content of S decreased significantly after the plasma treatment.

3. SERS MEASUREMENTS

**Raman Spectrum of Neurotransmitters**

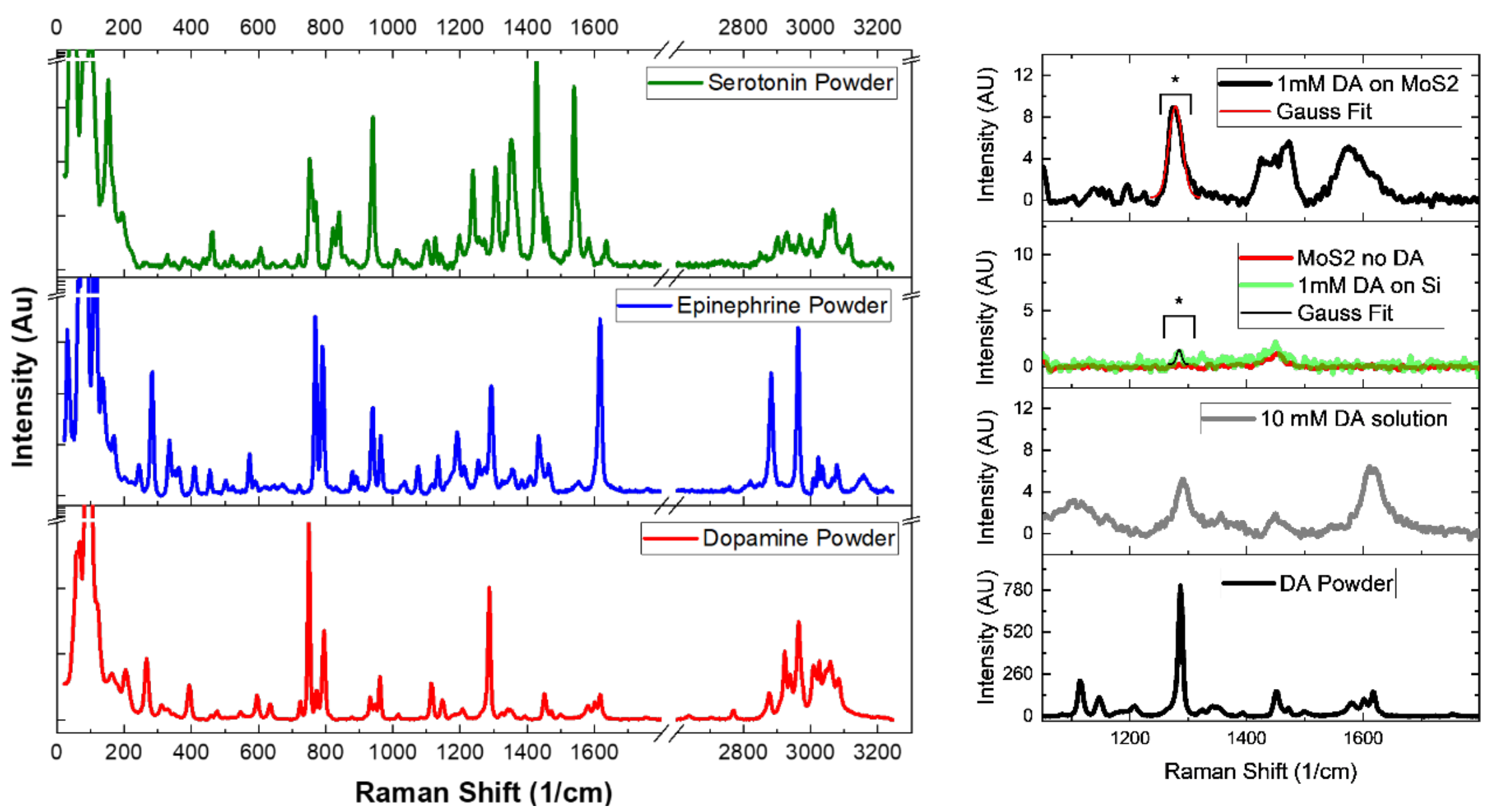

**Figure S6: (left)** Raman spectra of Dopamine (red), Epinephrine (blue) and Serotonin (green) powder. **(right)** Comparison of Raman spectra (from bottom to top) of Dopamine in powder form, 10 mM Dopamine in water solution, 1 mM Dopamine incubated for 2 hours on bare Si substrate, $MoS_2$ substrate (no DA), 1 mM Dopamine incubated for 2 hours on plasma-treated $MoS_2$ substrate.

**Enhancement factor calculation**

The theoretical diffraction-limited spot size for the 100x magnification objective lens was calculated using the formula $D = {}^{1.22\lambda}/_{NA}$ and the resulting laser spot area, $A = \pi(D/2)^2$, where λ represents the wavelength and the numerical aperture is NA.

Given our experimental parameters, laser spot area A was approximately $4.1\times10^{-9}$ cm$^2$.

The plasma treatment resulted in a net increase in defect density of $5.6\times10^{14}$ cm$^{-2}$. Assuming that catechol-mediated attachment requires two neighboring sulfur vacancies, the active-site density was estimated to be half of the plasma-induced defect density, yielding $2.8\times10^{14}$ cm$^{-2}$.

By considering the laser spot area (A), the number of active molecules contributing to the SERS signal ($N_{SERS}$) was determined to be $11.5\times10^5$.

Following incubation in a 1 mM dopamine solution for 2 hours, the plasma-treated sample exhibited an SERS peak intensity ($I_{SERS}$) of 0.15. As no discernible signal was observed in the reference sample (without plasma treatment $MoS_2$ incubated in DA solution), the maximum potential undetected Raman intensity ($I_{Ref,max}$) was conservatively defined using the (3σ) criterion (three times the standard deviation of the baseline).

The lower-bound SERS enhancement factor ($EF_{min}$) was calculated using the following relationship:

$$EF_{min} = \frac{I_{SERS}/N_{SERS}}{I_{Ref.,Max}/N_{Ref.}}$$

where, $I_{SERS}$ and $I_{Ref.,Max}$ represent the intensity of the Raman mode in the SERS regime and of the reference substrate, respectively: $N_{SERS}$ and $N_{Ref.}$ represent the number of molecules contributing to their respective signals. The actual number of molecules contributing to the reference signal is unknown and is expected to be lower than on the plasma-treated surface. Thus, we use the conservative assumption $N_{Ref} = N_{SERS}$. Under this assumption, a lower-bound EF of approximately 31.7±1.3 was determined.

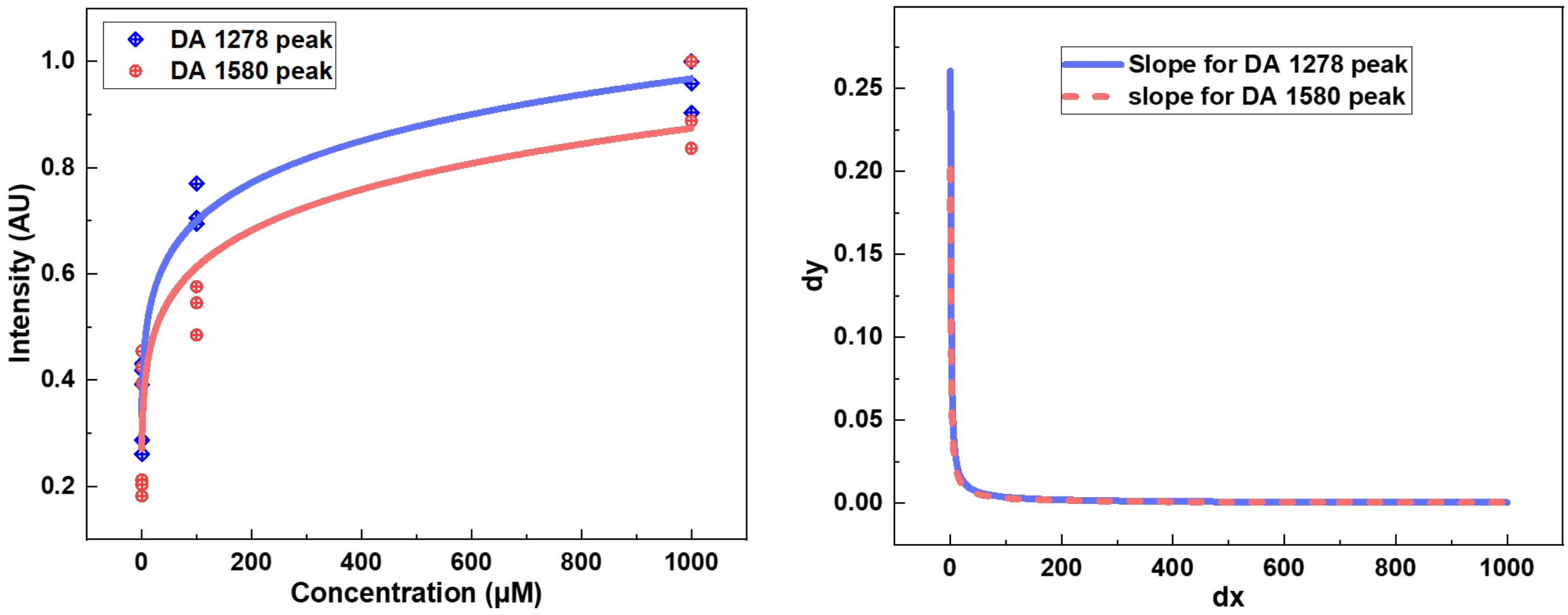

**Figure S7: (left)** Nonlinear DA SERS intensity dependence on concentration for the bands at 1278 $cm^{-1}$ and 1580 $cm^{-1}$ in the higher concentration range (0.5 μM–1 mM). **(right)** Derivative plots (dy/dx) of the nonlinear dependence, demonstrating similar slopes for both SERS bands.

Raman spectra of dopamine (DA), epinephrine (EP), and serotonin or 5-hydroxytryptamine (5-HT) powders are presented in Figure S6(left). The corresponding vibrational mode assignments for the Raman peaks of neurotransmitters are shown in Table S1.

The SERS intensity exhibited an approximately linear dependence on analyte concentration in the range of 0.5 nM to 1 μM (for DA $R^2$ = 0.95). At higher concentrations, the response gradually deviated from linearity and approached saturation, likely due to saturation of adsorption sites on the defect-engineered $MoS_2$ surface. Derivative analysis further confirmed a reduction in sensitivity at elevated concentrations, indicating the onset of adsorption-limited response (Figure S7).

Table S1: Vibrational mode assignment of powder DA, EP and 5-HT Raman spectra

| **Dopamine (DA)**[18–22] | | **Epinephrine (EP)**[23–26] | | **Serotonin (5-HT)**[27] | |
|---|---|---|---|---|---|
| This study | Peak assignment (literature) | This study | Peak assignment (literature) | This study | Peak assignment (literature) |
| 1115 | N-H twisting, Aliphatic chain C-H twisting, ring deformation, -OH twisting (1117) | 1079 | CCH bending (1081) | 1100 | CH rocking and NHCH2 rocking (1101) |
| 1148 | Ring Deformation, -OH bending (1147) | 1134 | Ring deformation, OH bending, chain twisting (1117) | 1126 | NHCH scissoring (1129) |
| 1207 | C-O stretching (1209) 76 | 1192 | 1177 stretching vibration of the C–O group | 1199 | OH rocking and ring CH scissoring (1200) |
| 1287 | C-O stretching, Ring breathing, CH rocking and twisting (1289) | 1293 | C-O stretching (1288) | 1238 | CH in plane rocking (1245) |
| 1345 | in-plane deformation modes of C–OH in catechol (1339) | 1363 | CH bending in benzene ring (1355) | 1305 | CH2 twist (1309) |
| 1393 | aliphatic chain CH2–CH2 bending (1381) | 1435 | aliphatic H-O-C-H bending; Chain twisting; -$CH_3$ bending (1431) | 1352 | Chain CH2 twist and CH rocking in pyrrole (1357) |
| 1452 | CH scissoring (1455) | 1467 | CH bending (1470) | 1428 | Chain CH2 wagging (1440) |
| 1473 | Aliphatic chain C-H bending (1473) | 1557 | bending of (C-H) bond in -NH-$CH_3$ (1548) | 1459 | -- |
| 1499 | C–C stretching mode in the aromatic (1481) | -- | C-C stretching (1599) | 1540 | Chaing CH2 scissoring and NH rocking (1546) |
| 1581 | C-C stretching/ $NH_2$ scissoring (1588) | 1617 | C=C stretching (1612) | 1582 | Chain CH2 scissoring (1583) |
| 1602 | Ring deformation and OH bending (1602) | | | 1635 | Benzene ring stretching (1635) |
| 1617 | Ring deformation and OH scissoring (1617) | | | | |

4. MACHINE LEARNING BASED MULTIVARIATE ANALYSIS

Our results demonstrate that defect-engineered $MoS_2$ is capable of detecting analytes containing the catechol functional group (DA and EP) while showing no significant response to other (5-HT). However, the proximity of the SERS peaks of the DA and EP poses a challenge in distinguishing them. For example, even though the most intense peak for DA appears at 1278 $cm^{-1}$, and epinephrine at 1284 $cm^{-1}$, both these peaks are broad and overlap. This fact raises concerns regarding the ability to differentiate between these neurotransmitters in real-world samples, where additional analytes and environmental factors would further complicate detection. To address this, we employed machine learning tools based on multivariate analysis, combining principal component analysis (PCA) and linear discriminant analysis (LDA), to evaluate whether the SERS spectra of dopamine and epinephrine could be effectively separated.

PCA is a statistical technique commonly used in spectral analysis to reduce the dimensionality of datasets while retaining the most significant features. PCA transforms the dataset into orthogonal components called principal components (PCs), which represent unique spectral patterns (loadings). The scores associated with these PCs reflect the correlation of each sample with these patterns. By plotting the scores of PCs, clustering patterns can be visualized, revealing spectral similarities or distinctions among samples [28,29]. This approach is particularly effective for analyzing complex datasets in Raman spectroscopy, where spectral features can vary subtly between analytes [29–32].

Given the similarity of the major SERS peaks of DA and EP, we hypothesized that the overall spectral patterns, including peak intensity and shape, would provide sufficient variance for differentiation. To test this, we performed PCA on background-subtracted and Si-normalized SERS spectra.

We began by analyzing 10 spectra each of dopamine and epinephrine at concentrations of 1 µM and 50 nM. The PCA score plot for high concentrations (DA 1 µM and EP 1 µM) and for low concentrations (DA 50 nM and EP 50 nM) showed clear clustering and separation along the PC1 axis (Figure S8a and S8b).

To further test the method's utility, we performed PCA on spectra from a mixed dataset, combining 20 spectra of DA (1 µM and 50 nM) and 20 spectra of EP (1 µM and 50 nM). The scatter plot showed distinct clustering into four regions based on both analyte type and concentration, with ellipses representing a 95% confidence level (Figure S8c). This finding confirmed that, despite the overlap in individual SERS peaks, the spectral patterns contain sufficient variance to differentiate between analytes and their concentrations.

Next, we performed LDA on the first four PCs (results of Figure S8c) of the mixed-concentration dataset (40 spectra: 20 each of dopamine and epinephrine at 1 µM and 50 nM). While PCA provides an unsupervised dimensionality reduction approach, LDA introduces a supervised classification method to further enhance differentiation. LDA identifies a hyperplane that maximally separates classes (e.g., analytes) based on predefined labels, making it an ideal follow-up to PCA for classification tasks. To evaluate the robustness of the classification model, 5-fold cross-validation was applied during the LDA analysis. Figure S8d shows the LDA projection, with clear separation of the two neurotransmitters based on their class labels. Using a custom Python-based machine learning algorithm, 30% of the data was used for training, and 70% was used for testing, further validating the model's predictive performance. This approach achieved a classification accuracy of 96%, as further validated by the confusion matrix (Figure S8d-bottom panel). The confusion matrix demonstrates that the LDA model applied to PCA scores accurately classified 27 out of 28 test samples, with only one epinephrine sample misclassified as dopamine.

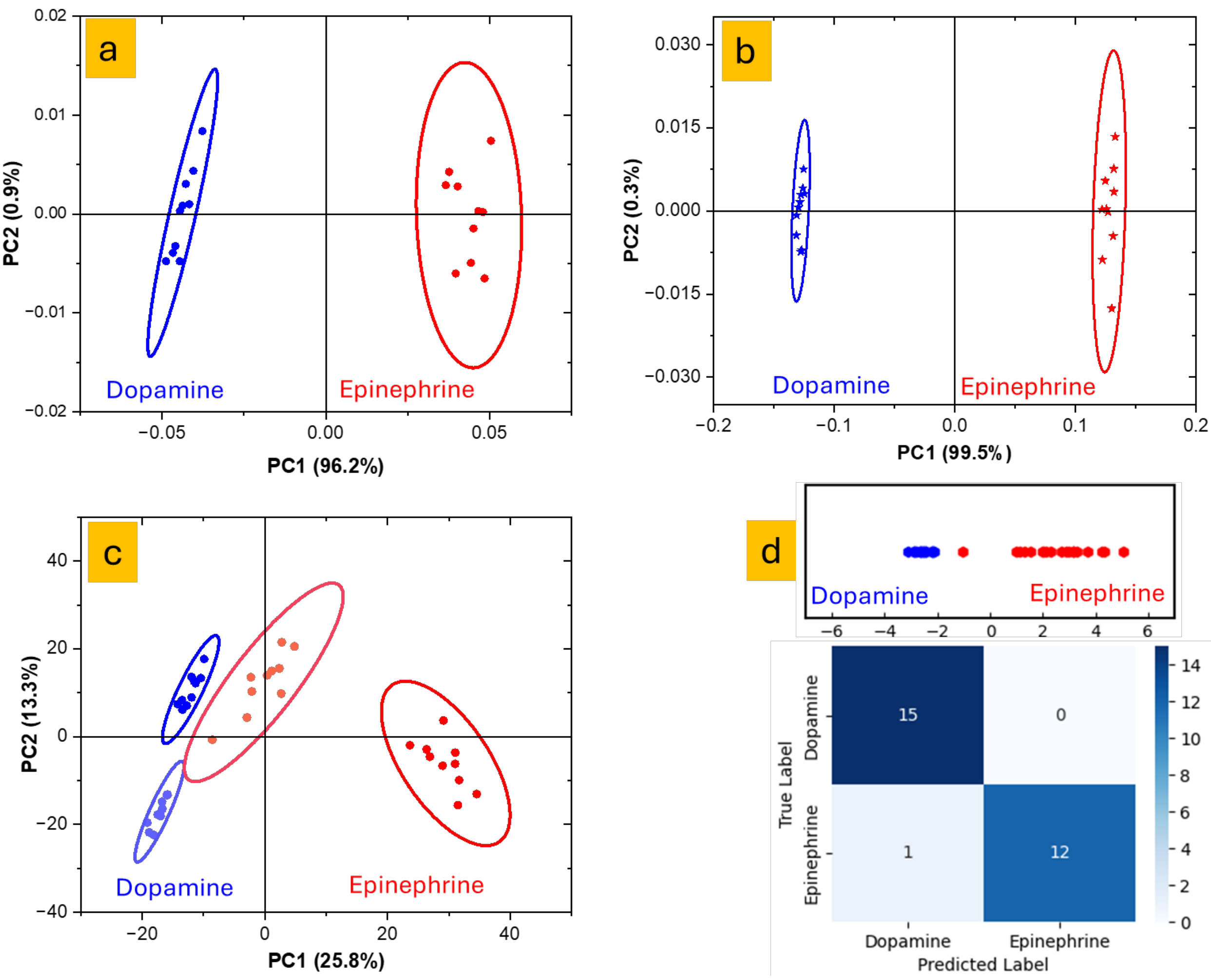

**Figure S8:** Score plot of PC1 vs. PC2 obtained from PCA analysis of dopamine and epinephrine at 1 µM concentration (a) and 50 nM concentration (b), with each dataset containing 10 spectra. (c) Combined score plot of PC1 vs. PC2 from PCA analysis of dopamine and epinephrine, encompassing two concentrations (1 µM and 50 nM), with 10 spectra per class at each concentration. (d) LDA performed on the PCA scores shown in (c) to differentiate dopamine and epinephrine based on their spectral characteristics, (bottom panel represents the confusion matrix). All ellipses represent a 95% confidence interval.

The effect of the number of PCs on LDA accuracy across different cross-validation folds is summarized in Figure S9. Utilizing the first four PCs produced the highest classification accuracy, while adding more PCs decreased accuracy, likely due to noise contributions. This highlights the significance of optimizing the number of PCs to balance signal variance and noise in multivariate analysis. A 5-fold cross-validation method was implemented to evaluate LDA's performance on the PCA-transformed data. Classification accuracy increased with the addition of principal components, achieving optimal performance with the first four PCs, as depicted in Figure S9a. After the fourth PC, overall accuracy declined, likely due to the addition of lower-variance components that could capture noise instead of meaningful signals. Figure S9b demonstrates how LDA classification accuracy depends on the number of principal components and the folds used in the cross-validation process. The solid olive curve in Figure S9b represents results from using the first four PCs, showing consistent 100% accuracy across all fold values, except for the 3$^{rd}$-fold cross-validation, which showed an accuracy of 88%. Conversely, using only the first PC led to a significant accuracy drop, down to as low as 50% (solid black curve). Other curves in Figure S9b reveal various levels of accuracy depending on the combination of PCs and folds applied, highlighting the importance of selecting an optimal number of components for effective classification performance.

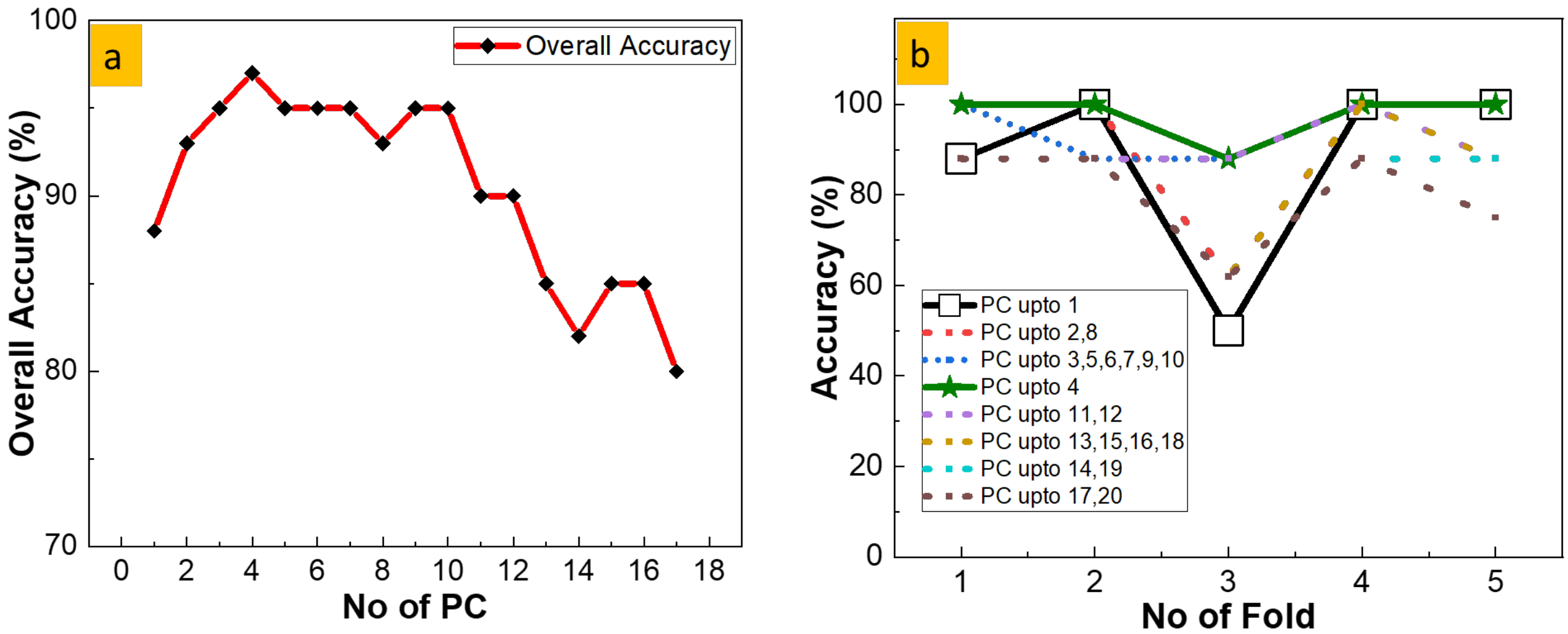

**Figure S9:** (a) Overall classification accuracy of DA and EP as a function of the number of principal components (PCs) used for LDA analysis, evaluated using a 5-fold cross-validation method. The accuracy varies depending on the number of PCs included in the analysis. (b) Classification accuracy across different folds in the 5-fold cross-validation method for varying subsets of PCs. Each line represents the number of PCs used, with only one line shown for cases where the total accuracy is identical across multiple PC subsets. The optimal performance is highlighted for PCs up to 4 (line in olive).

Figure S10 illustrates the confusion matrices that summarize the performance of LDA implemented with a 5-fold cross-validation method using 4 PCs. The analysis utilized a total of 40 Raman spectra, divided into 20 spectra for dopamine and 20 for epinephrine. In each fold, 30% of the data was served for training while 70% was dedicated to testing. The overall confusion matrix presented in Figure S10a reveals that only one epinephrine sample was incorrectly classified as dopamine, emphasizing the LDA model's high classification accuracy. Figure S10 b–f showcase the confusion matrices from each individual fold, with only Fold 3 reflecting a single misclassification, where one epinephrine sample was wrongly categorized as dopamine. All other folds attained perfect classification, effectively differentiating between dopamine and epinephrine.

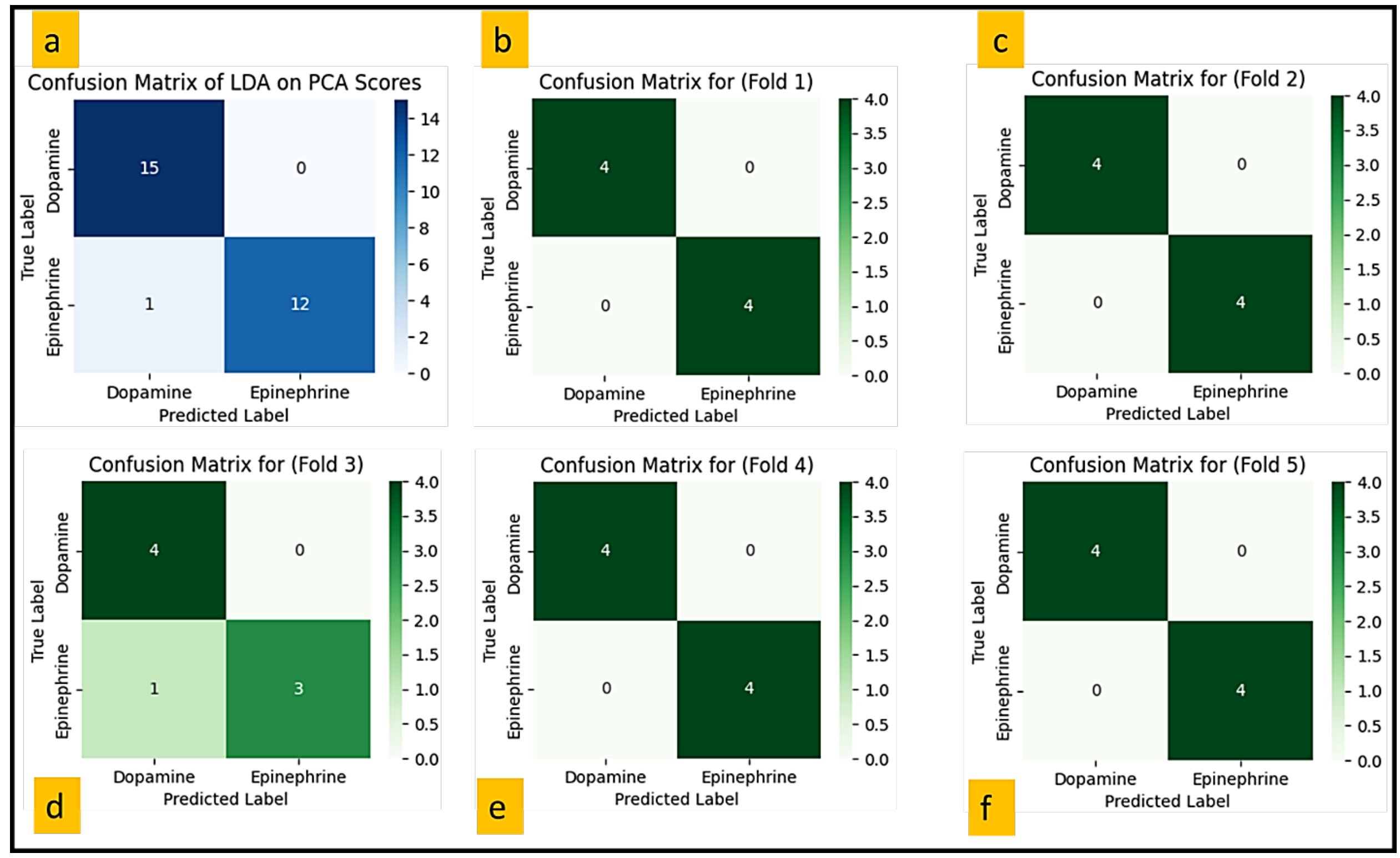

**Figure S10:** (a) Confusion matrix of the LDA analysis using PCA scores, where 70% of the data was used for testing. The matrix demonstrates the classification performance of dopamine and epinephrine. (b–f) Confusion matrix for each fold in a 5-fold cross-validation analysis performed using the first four principal components (PCs). Each fold illustrates the classification accuracy and consistency in distinguishing dopamine and epinephrine across all validation sets.

Based on these understanding of incorporating machine learning into multivariate analysis (PCA–LDA), we performed this analysis framework on all samples across four concentration levels (1 µM to 0.5 nM), together with control and negative control spectra. The final classification model achieved 100% accuracy when trained on 40% of the data and tested on the remaining 60%, as detailed in the main paper. It is also noteworthy that during model development, when 30% of the data was used for training, the accuracy remained as high as 99%, as presented in Figure S11.

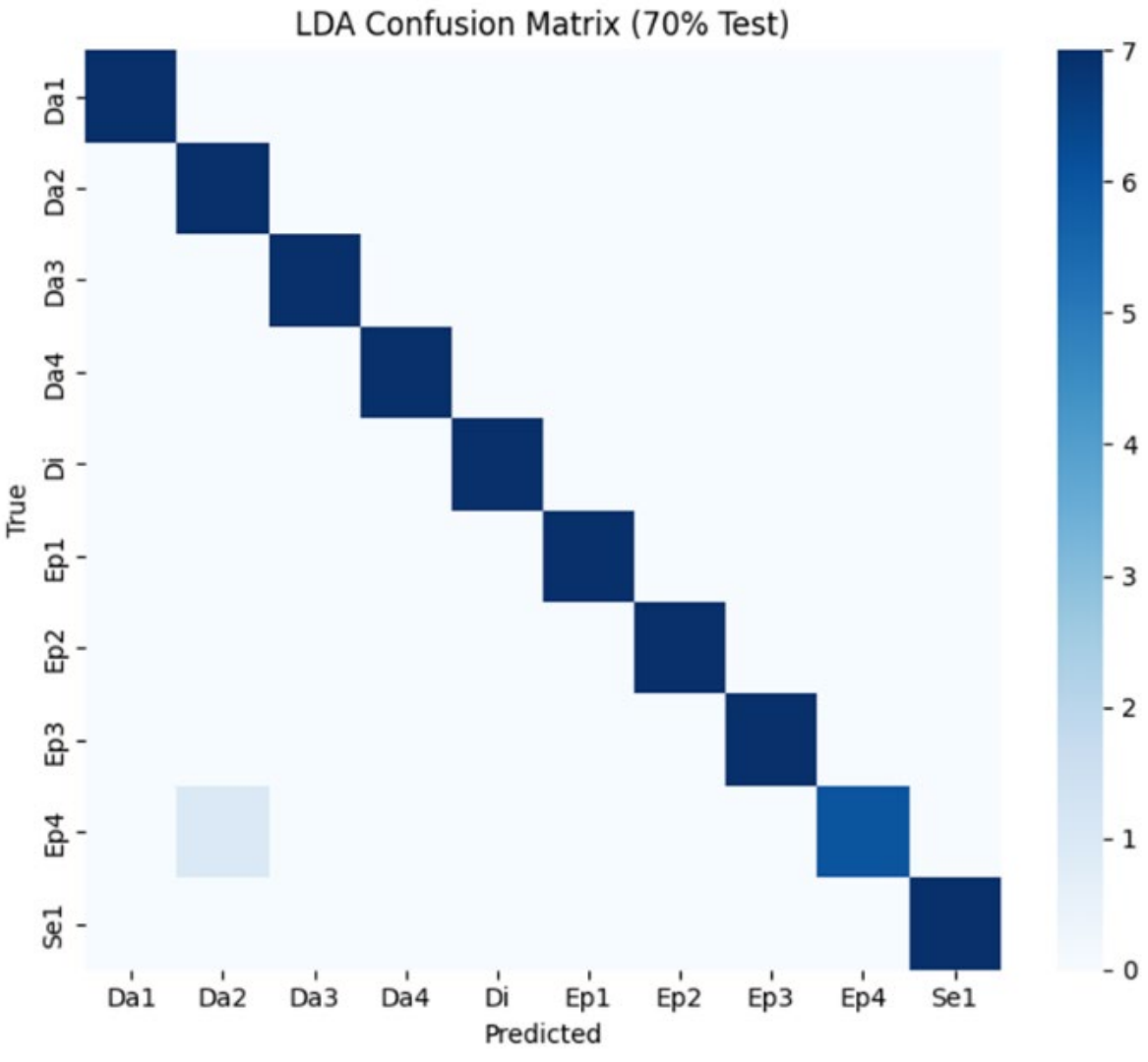

**Figure S11:** Confusion matrix of the LDA analysis using PCA scores, where 30% of the data was used for training and remaining 70% was used for testing (99% accuracy).

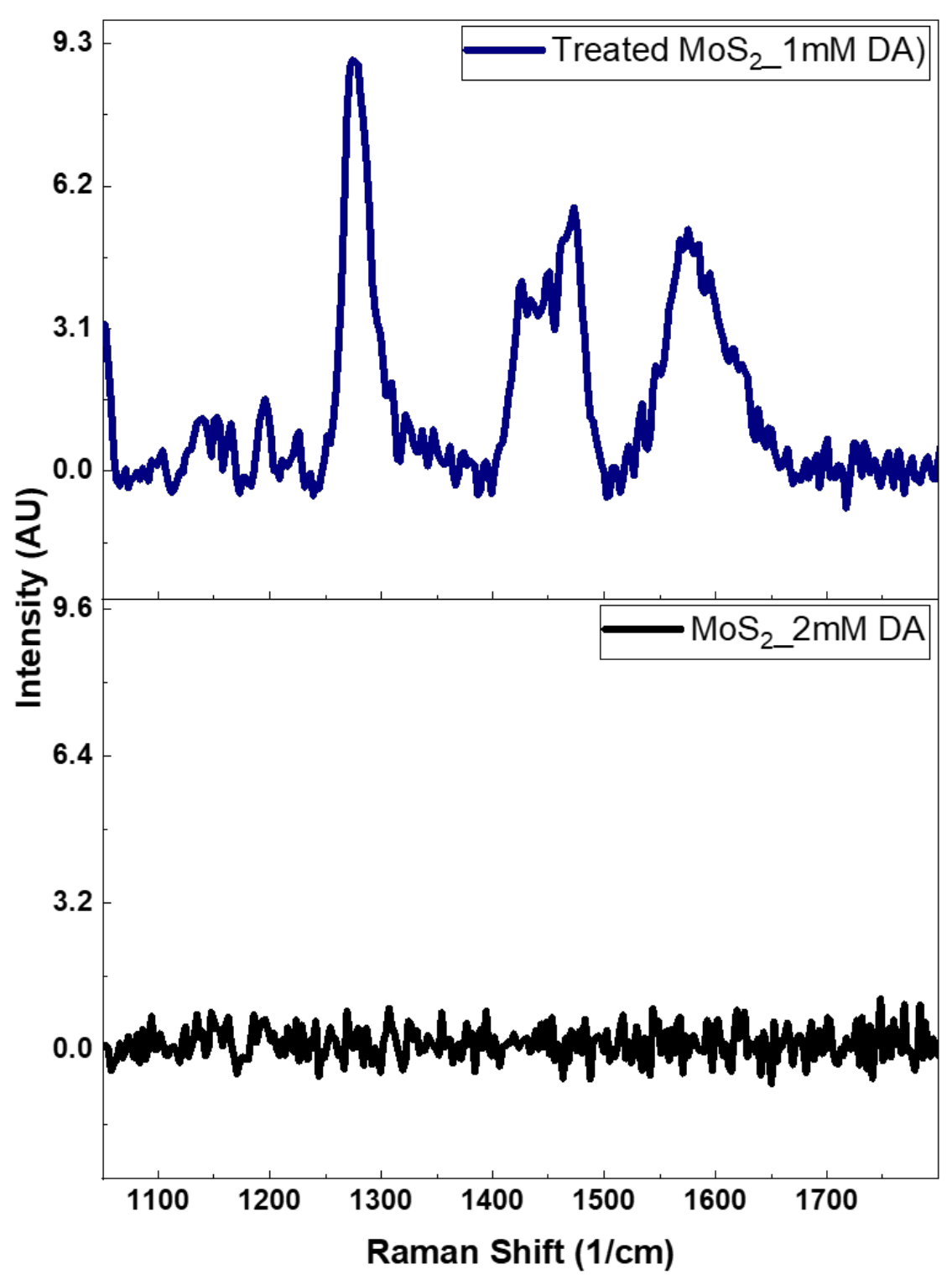

**Figure S12:** Raman spectra of DA on pristine and plasma-treated $MoS_2$ under identical incubation conditions, showing enhanced signal intensity from defect-engineered $MoS_2$.

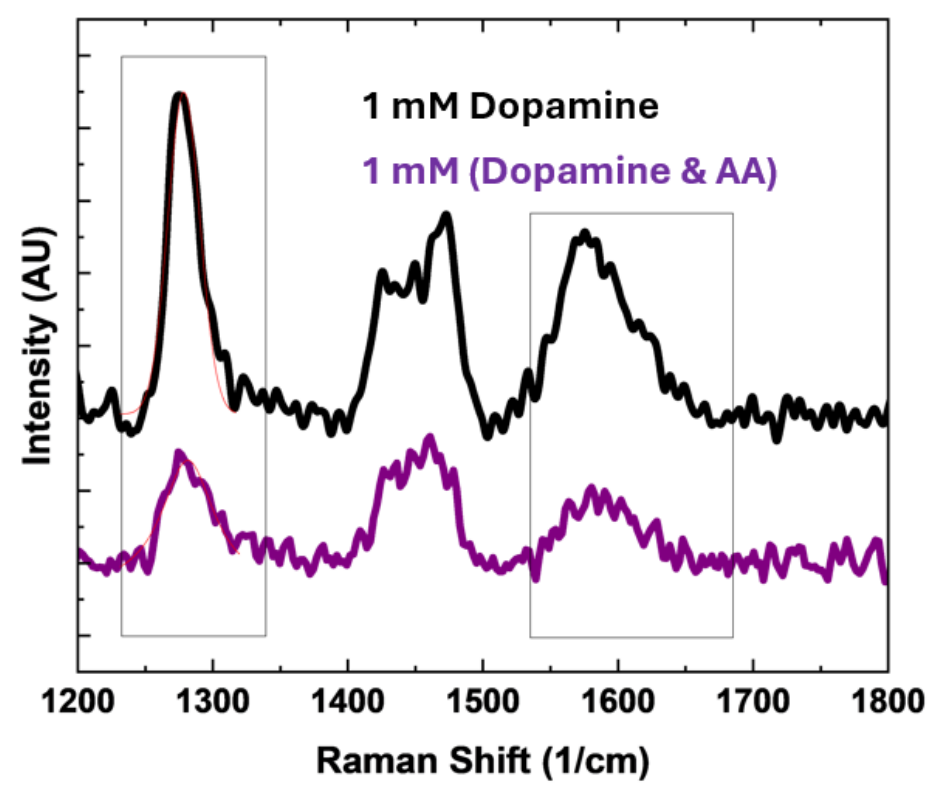

**Figure S13:** Comparison of detection of 1mM DA in DI and 1mM DA & 1 mM Ascorbic Acid (AA). We observed a drop in intensity and broadening of dopamine spectral lines, but the characteristic Raman peaks of dopamine remain clearly distinguishable, indicating that the $MoS_2$ substrate retains chemical sensitivity to dopamine even in the presence of a competing redox-active molecule.

5. ADDITIONAL SUPPLEMENTAL INFORMATION

**Python code for the LDA model**

LDA model python code.py

```
1    import pandas as pd
2    import numpy as np
3    from sklearn.discriminant_analysis import LinearDiscriminantAnalysis as LDA
4    from sklearn.model_selection import train_test_split
5    from sklearn.metrics import accuracy_score, confusion_matrix
6    import matplotlib.pyplot as plt
7    import seaborn as sns
8    from sklearn.model_selection import StratifiedKFold,
cross_val_score,cross_val_predict
9    from sklearn.metrics import accuracy_score, confusion_matrix, classification_report
10
11   # Step 1: Load Data from a CSV file
12   CSV_PATH = r"C:\file\path\to\PCAresults.csv"
13   PC_COLUMNS = ['PC1', 'PC2']    # change if needed (must exist in the CSV)
14
15   # Load data and basic cleaning
16
17   df = pd.read_csv(CSV_PATH)
18
19   # Clean Class text (optional)
20   df['Class'] = df['Class'].astype(str).str.strip().str.capitalize()
21
22   # Keep only PCs that exist in the file
23   pc_cols = [c for c in PC_COLUMNS if c in df.columns]
24   if len(pc_cols) == 0:
25       raise ValueError(f"None of {PC_COLUMNS} found in CSV columns:
{list(df.columns)}")
26
27   X = df[pc_cols].copy()
28   y = df['Class'].copy()
29
30   # Replace inf with NaN, drop rows with NaN
31   X = X.replace([np.inf, -np.inf], np.nan).dropna()
32   y = y.loc[X.index]   # align y to cleaned X
33
34   # Step 2) Stratified 30/70 train/test split (UNSEEN test for unbiased eval)
35
36   X_train, X_test, y_train, y_test = train_test_split(
37       X, y,
38       test_size=0.6,          # 60% test, 40% train (as requested)
39       stratify=y,
40       random_state=42
41   )
42
43   # Determine LDA components from TRAINING ONLY to avoid leakage
44   n_comp = min(2, len(np.unique(y_train)) - 1, X_train.shape[1]) # for multicomponent
LD1 vs LD2
45   lda = LDA(n_components=n_comp)
46   #n_comp = 1       # for LD1 onoly
47   #lda = LDA(n_components=1)
48
49   print(f"\nUsing {n_comp} LDA component(s) "
50         f"(min(2, #classes-1={len(np.unique(y_train))-1},
#features={X_train.shape[1]})).")
51
52   # Step 3) 4-fold CV on TRAINING set only (no leakage)
53
54   cv = StratifiedKFold(n_splits=4, shuffle=True, random_state=42)
55
56   cv_scores = cross_val_score(lda, X_train, y_train, cv=cv, scoring='accuracy')
57   print(f"\nTraining 5-fold CV accuracies: {cv_scores}")
```

```
58   print(f"Training 5-fold CV mean: {cv_scores.mean():.3f} ± {cv_scores.std():.3f}")
59
60   # Optional: Out-of-fold predictions on training set for a training report
61   y_train_oof = cross_val_predict(lda, X_train, y_train, cv=cv)
62   print("\nTraining (OOF) classification report:\n",
63         classification_report(y_train, y_train_oof))
64
65   # Step 4) Final fit on TRAINING set and unbiased TEST evaluation
66
67   lda.fit(X_train, y_train)
68   y_test_pred = lda.predict(X_test)
69
70   print("\n=== UNBIASED TEST RESULTS (70% hold-out) ===")
71   print("Test accuracy:", f"{accuracy_score(y_test, y_test_pred):.3f}")
72   print("\nTest classification report:\n", classification_report(y_test, y_test_pred))
73
74   # Confusion matrix on TEST
75   cm_test = confusion_matrix(y_test, y_test_pred, labels=np.unique(y))
76   plt.figure(figsize=(7, 6))
77   sns.heatmap(cm_test, annot=False, cmap='Blues',
78               xticklabels=np.unique(y), yticklabels=np.unique(y))
79   plt.xlabel('Predicted'); plt.ylabel('True'); plt.title('LDA Confusion Matrix (70%
Test)')
80   plt.tight_layout(); plt.show()
81
82   # Step 5) Visualization: LD1 vs LD2 (or 1D) with TRAIN vs TEST
83
84   Xtr_lda = lda.transform(X_train)
85   Xte_lda = lda.transform(X_test)
86
87   classes = np.unique(y)
88   palette = plt.get_cmap('tab20')
89   color_map = {cls: palette(i % 20) for i, cls in enumerate(classes)}
90
91   if n_comp >= 2:
92       # 2D scatter
93       plt.figure(figsize=(7, 5))
94       for cls in classes:
95           # Train points
96           pts_tr = Xtr_lda[y_train.values == cls]
97           plt.scatter(pts_tr[:, 0], pts_tr[:, 1],
98                       s=28, marker='o', color=color_map[cls], alpha=0.8, label=f"{cls}
(train)")
99           # Test points
100           pts_te = Xte_lda[y_test.values == cls]
101          if pts_te.size > 0:
102              plt.scatter(pts_te[:, 0], pts_te[:, 1],
103                          s=38, marker='^', facecolors='none',
edgecolors=color_map[cls], linewidths=1.2,
104                          label=f"{cls} (test)")
105      plt.xlabel('LD1', weight='bold'); plt.ylabel('LD2', weight='bold')
106      plt.title('LDA (train vs test)', weight='bold')
107      # Reduce legend clutter: one entry per class (train & test combined)
108      handles, labels = plt.gca().get_legend_handles_labels()
109      # Keep unique labels order
110      uniq = dict()
111      for h, lab in zip(handles, labels):
112          if lab not in uniq:
113              uniq[lab] = h
114      plt.legend(uniq.values(), uniq.keys(), ncol=2, fontsize=8, frameon=False)
115      plt.tight_layout(); plt.savefig('lda_ld1_ld2_train_vs_test.png', dpi=300,
transparent=True)
116      plt.show()
```

```
117  else:
118      # 1D strip (if only LD1 is available)
119      plt.figure(figsize=(7, 2.8))
120      # Train
121      for cls in classes:
122          xi = Xtr_lda[y_train.values == cls, 0]
123          jitter = (np.random.rand(xi.size) - 0.5) * 0.05
124          plt.scatter(xi, jitter, s=26, marker='o', color=color_map[cls], alpha=0.8,
label=f"{cls} (train)")
125      # Test
126      for cls in classes:
127          xi = Xte_lda[y_test.values == cls, 0]
128          jitter = (np.random.rand(xi.size) - 0.5) * 0.05
129          if xi.size > 0:
130              plt.scatter(xi, jitter, s=36, marker='^', facecolors='none',
edgecolors=color_map[cls],
131                          linewidths=1.2, label=f"{cls} (test)")
132      plt.xlabel('LD1', weight='bold'); plt.yticks([])
133      plt.title('LDA (train vs test)', weight='bold')
134      handles, labels = plt.gca().get_legend_handles_labels()
135      uniq = dict()
136      for h, lab in zip(handles, labels):
137          if lab not in uniq:
138              uniq[lab] = h
139      plt.legend(uniq.values(), uniq.keys(), ncol=2, fontsize=8, frameon=False)
140      plt.tight_layout(); plt.savefig('lda_ld1_train_vs_test.png', dpi=300,
transparent=True)
141      plt.show()
142
143  # ============================================================
```

**References:**

1 M. A. R. Khan, S. Kalkar, B. Khader, O. O. Ayodele, A. Prokofjevs and T. Ignatova, *Nanoscale*, 2025, **17**, 19870–19881.

2 F. Bussolotti, K. E. J. Goh, J. Yang, H. Kawai and C. P. Y. Wong, *ACS Nano*, 2021, **15**, 2686–2697.

3 F. Bussolotti, J. Chai, M. Yang, H. Kawai, Z. Zhang, S. Wang, S. L. Wong, C. Manzano, Y. Huang, D. Chi and K. E. J. Goh, *RSC Adv.*, 2018, **8**, 7744–7752.

4 Y. Zhang, T. R. Chang, B. Zhou, Y. T. Cui, H. Yan, Z. Liu, F. Schmitt, J. Lee, R. Moore, Y. Chen, H. Lin, H. T. Jeng, S. K. Mo, Z. Hussain, A. Bansil and Z. X. Shen, *Nat. Nanotechnol.*, 2014, **9**, 111–115.

5 Y. Kim, H. Bark, G. H. Ryu, Z. Lee and C. Lee, *Journal of Physics: Condensed Matter*, 2016, **28**, 184002.

6 C. A. Papageorgopoulos and W. Jaegermann, *Surf. Sci.*, 1995, **338**, 83–93.

7 T. Grünleitner, A. Henning, M. Bissolo, M. Zengerle, L. Gregoratti, M. Amati, P. Zeller, J. Eichhorn, A. V. Stier, A. W. Holleitner, J. J. Finley and I. D. Sharp, *ACS Nano*, 2022, **16**, 20364–20375.

8 G. Deokar, D. Vignaud, R. Arenal, P. Louette and J. F. Colomer, *Nanotechnology*, DOI:10.1088/0957-4484/27/7/075604.

9 J. H. Bihn, J. Park and Y. C. Kang, *Journal of the Korean Physical Society*, 2011, **58**, 509–514.

10 T. Weber, J. C. Muijsers, J. H. M. C. Van Wolput, C. P. J. Verhagen and J. W. Niemantsverdriet, *Journal of Physical Chemistry*, 1996, **100**, 14144–14150.

11 F. Bussolotti, K. E. J. Goh, J. Yang, H. Kawai and C. P. Y. Wong, *ACS Nano*, 2021, **15**, 2686–2697.

12 A. V Shchukarev and D. V Korolkov, *Central European Science Journals Central European Journal of Chemistry XPS Study of Group IA Carbonates*, .

13 Y. Zhu, J. Lim, Z. Zhang, Y. Wang, S. Sarkar, H. Ramsden, Y. Li, H. Yan, D. Phuyal, N. Gauriot, A. Rao, R. L. Z. Hoye, G. Eda and M. Chhowalla, *ACS Nano*, 2023, **17**, 13545–13553.

14 M. Rajput, S. K. Mallik, S. Chatterjee, A. Shukla, S. Hwang, S. Sahoo, G. V. P. Kumar and A. Rahman, *Commun. Mater.*, DOI:10.1038/s43246-024-00632-y.

15 J. H. Scofield, *TID-4500, UC-34 Physics THEORETICAL PHOTOIONIZATION CROSS SECTIONS FROM I TO 1500 keV*, 1973.

16 A. V Kolobov and J. Tominaga, *Springer Series in Materials Science 239 Two-Dimensional Transition-Metal Dichalcogenides*, .

17 L. Gao, Q. Liao, X. Zhang, X. Liu, L. Gu, B. Liu, J. Du, Y. Ou, J. Xiao, Z. Kang, Z. Zhang and Y. Zhang, *Advanced Materials*, 2020, **32**, 1906646.

18 I. Badillo-Ramírez, J. M. Saniger, J. Popp and D. Cialla-May, *Physical Chemistry Chemical Physics*, 2021, **23**, 12158–12170.

19 M. Kim, Y. S. Choi and D. H. Jeong, *RSC Adv.*, 2024, **14**, 14214–14220.

20 A. Hariharan, R. Kurnoothala, S. K. Chinthakayala, K. C. Vishnubhatla and P. Vadlamudi, *Spectrochim. Acta A Mol. Biomol. Spectrosc.*, DOI:10.1016/j.saa.2021.119962.

21 J. D. Ciubuc, K. E. Bennet, C. Qiu, M. Alonzo, W. G. Durrer and F. S. Manciu, *Biosensors (Basel).*, DOI:10.3390/bios7040043.

22 A. Michałowska, K. Jędrzejewski and A. Kudelski, *Materials*, DOI:10.3390/ma15175972.

23 A. Hariharan and P. Vadlamudi, *J. Mol. Struct.*, DOI:10.1016/j.molstruc.2021.131163.

24 O. E. Eremina, N. R. Yarenkov, O. O. Kapitanova, A. S. Zelenetskaya, E. A. Smirnov, T. N. Shekhovtsova, E. A. Goodilin and I. A. Veselova, *Microchimica Acta*, DOI:10.1007/s00604-022-05247-z.

25 C. Shende, W. Smith, C. Brouillette and S. Farquharson, *Pharmaceutics*, 2014, **6**, 651–662.

26 R. L. Araújo, J. X. Lima Neto, C. A. Barboza, J. I. N. Oliveira, R. M. Tromer, J. M. Henriques and U. L. Fulco, *J. Appl. Phys.*, DOI:10.1063/5.0054383.

27 P. Song, X. Guo, Y. Pan, Y. Wen, Z. Zhang and H. Yang, *Journal of Electroanalytical Chemistry*, 2013, **688**, 384–391.

28 J. Chi, Y. Ma, F. L. Weng, H. Thiessen-Philbrook, C. R. Parikh and H. Du, *J. Biophotonics*, 2017, **10**, 1743–1755.

29 M. Marro, A. Taubes, P. Villoslada and D. Petrov, *https://doi.org/10.1117/12.921358*, 2012, **8427**, 228–234.

30 B. Sharma, P. Bugga, L. R. Madison, A. I. Henry, M. G. Blaber, N. G. Greeneltch, N. Chiang, M. Mrksich, G. C. Schatz and R. P. Van Duyne, *J. Am. Chem. Soc.*, 2016, **138**, 13952–13959.

31 W. R. Premasiri, Y. Gebregziabher and L. D. Ziegler, *http://dx.doi.org/10.1366/10-06173*, 2011, **65**, 493–499.

32 A. K. Boardman, W. S. Wong, W. R. Premasiri, L. D. Ziegler, J. C. Lee, M. Miljkovic, C. M. Klapperich, A. Sharon and A. F. Sauer-Budge, *Anal. Chem.*, 2016, **88**, 8026–8035.